\documentclass[
reprint,
nofootinbib,
amsmath,amssymb,aps,pra,
onecolumn,notitlepage,
10pt,
superscriptaddress,
longbibliography
]{revtex4-1}

\usepackage[utf8]{inputenc}

\usepackage{natbib}

\usepackage{amsmath}
\usepackage{graphicx}   
\usepackage{dsfont}
\usepackage[colorinlistoftodos]{todonotes}
\usepackage[colorlinks=true, allcolors=blue]{hyperref}  
\usepackage{braket}     
\usepackage{algpseudocode}
\usepackage{tcolorbox}
\usepackage{algorithm}

\usepackage{mathtools}
\DeclarePairedDelimiter{\ceil}{\lceil}{\rceil}
\DeclarePairedDelimiter{\floor}{\lfloor}{\rfloor}

\usepackage{soul}

\newcommand{\la}{\langle}
\newcommand{\ra}{\rangle}
\newcommand{\kb}{\ra\la}

\newcommand{\half}{\frac{1}{2}}

\newcommand{\llp}{|l \ra\la l'|}
\newcommand{\lpl}{|l' \ra\la l|}

\newcommand{\lkbl}{|l \ra\la l|}

\newcommand{\sx}{\hat \sigma_x}
\newcommand{\sy}{\hat \sigma_y}
\newcommand{\sz}{\hat \sigma_z}

\newcommand{\Sx}{\hat S_x}
\newcommand{\Sy}{\hat S_y}
\newcommand{\Sz}{\hat S_z}
\newcommand{\Sint}{\hat S_z^{(i)} \hat S_z^{(j)}}

\newcommand{\inttwo}{\hat{a}_i ^\dag \hat{a}_{i+1} + \textrm{h.c.}}

\newcommand{\intelcali}{Intel Labs, Santa Clara, California 95054, USA}
\newcommand{\inteloregon}{Intel Labs, Hillsboro, Oregon 97124, USA}
\newcommand{\utchem}{Department  of  Chemistry,  University  of  Toronto,  Toronto,  Ontario  M5G 1Z8,  Canada}
\newcommand{\utcomp}{Department  of  Computer Science,  University  of  Toronto,  Toronto,  Ontario  M5S 2E4,  Canada}
\newcommand{\vectorinst}{Vector  Institute  for  Artificial  Intelligence,  Toronto,  Ontario  M5S  1M1,  Canada}
\newcommand{\cifar}{Canadian  Institute  for  Advanced  Research,  Toronto,  Ontario  M5G  1Z8,  Canada}

\begin{document}

\title{Resource-efficient digital quantum simulation of $d$-level systems for photonic, vibrational, and spin-$s$ Hamiltonians}




\author{Nicolas P. D. Sawaya}
\email{nicolas.sawaya@intel.com}
\affiliation{\intelcali}
\author{Tim Menke}
\affiliation{Department of Physics, Harvard University, Cambridge, MA 02138, USA}
\affiliation{Research Laboratory of Electronics, Massachusetts Institute of Technology, Cambridge, MA 02139, USA}
\affiliation{Department of Physics, Massachusetts Institute of Technology, Cambridge, MA 02139, USA}
\author{Thi Ha Kyaw}
\affiliation{\utcomp}
\affiliation{\utchem}\
\author{Sonika Johri}
\affiliation{\inteloregon}
\author{Al\'an Aspuru-Guzik}
\affiliation{\utcomp}
\affiliation{\utchem}
\affiliation{\vectorinst}
\affiliation{\cifar}
\author{Gian Giacomo Guerreschi}
\email{gian.giacomo.guerreschi@intel.com}
\affiliation{\intelcali}


\begin{abstract}

Simulation of quantum systems is expected to be one of the most important applications of quantum computing, with much of the theoretical work so far having focused on fermionic and spin-$\half$ systems. Here, we instead consider encodings of $d$-level (\textit{i.e.} qudit) quantum operators into multi-qubit operators, studying resource requirements for approximating operator exponentials by Trotterization. We primarily focus on spin-$s$ and truncated bosonic operators in second quantization, observing desirable properties for approaches based on the Gray code, which to our knowledge has not been used in this context previously. After outlining a methodology for implementing an arbitrary encoding,  we investigate the interplay between Hamming distances, sparsity patterns, bosonic truncation, and other properties of local operators. Finally, we obtain resource counts for five common Hamiltonian classes used in physics and chemistry, while modeling the possibility of converting between encodings within a Trotter step. The most efficient encoding choice is heavily dependent on the application and highly sensitive to $d$, although clear trends are present. These operation count reductions are relevant for running algorithms on near-term quantum hardware because the savings effectively decrease the required circuit depth. Results and procedures outlined in this work may be useful for simulating a broad class of Hamiltonians on qubit-based digital quantum computers.

\end{abstract}

\maketitle





\section{Introduction}
\label{sec:introduction}


%
\begin{figure}
    \centering
    \includegraphics[width=1.0\textwidth]{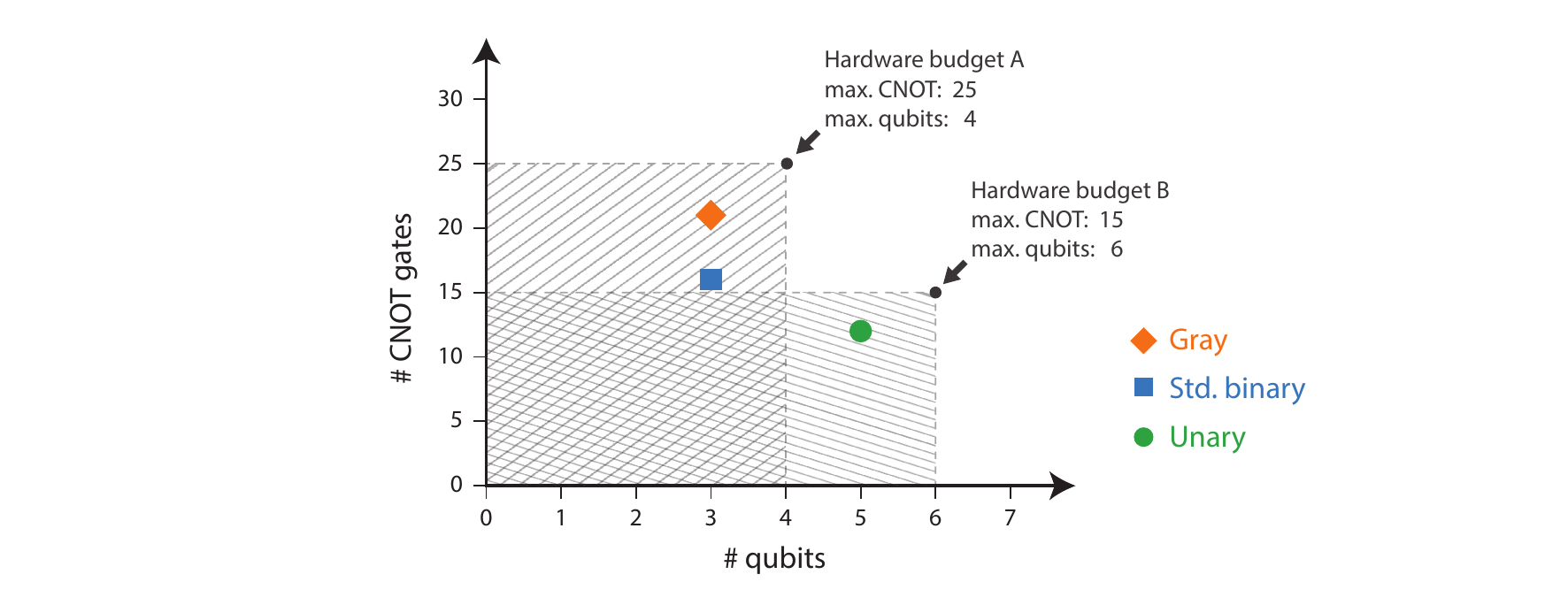}
    \caption{Especially for near-term noisy quantum hardware, gate counts and qubit counts will be limited. In principle, these constraints can be used to approximate a \textit{hardware budget} for a set of hardware and a particular Hamiltonian simulation problem. For example, if one wants to simulate a collection of $N$ bosons on a small quantum computer, the decoherence time and gate errors will constrain the allowed number of gates, while the total number of qubits will constrain the qubit count per boson. In this schematic, we show two arbitrary hardware budgets for Trotterizing the exponential of $\hat q^2$ for one boson with truncation $d=5$. In device A, both the Gray and standard binary encodings are satisfactory, but the unary code requires too many qubits. However, because device B allows for more qubits but fewer operations, the unary code is sufficient while the former two encodings require too many operations. This highlights the need for considering multiple encodings, as an encoding that is best for one type of hardware is not necessarily universally superior.  
    }
    \label{fig:hardware_budget}
\end{figure}
%

%
\begin{figure}
    \centering
    \includegraphics[width=1.0\textwidth]{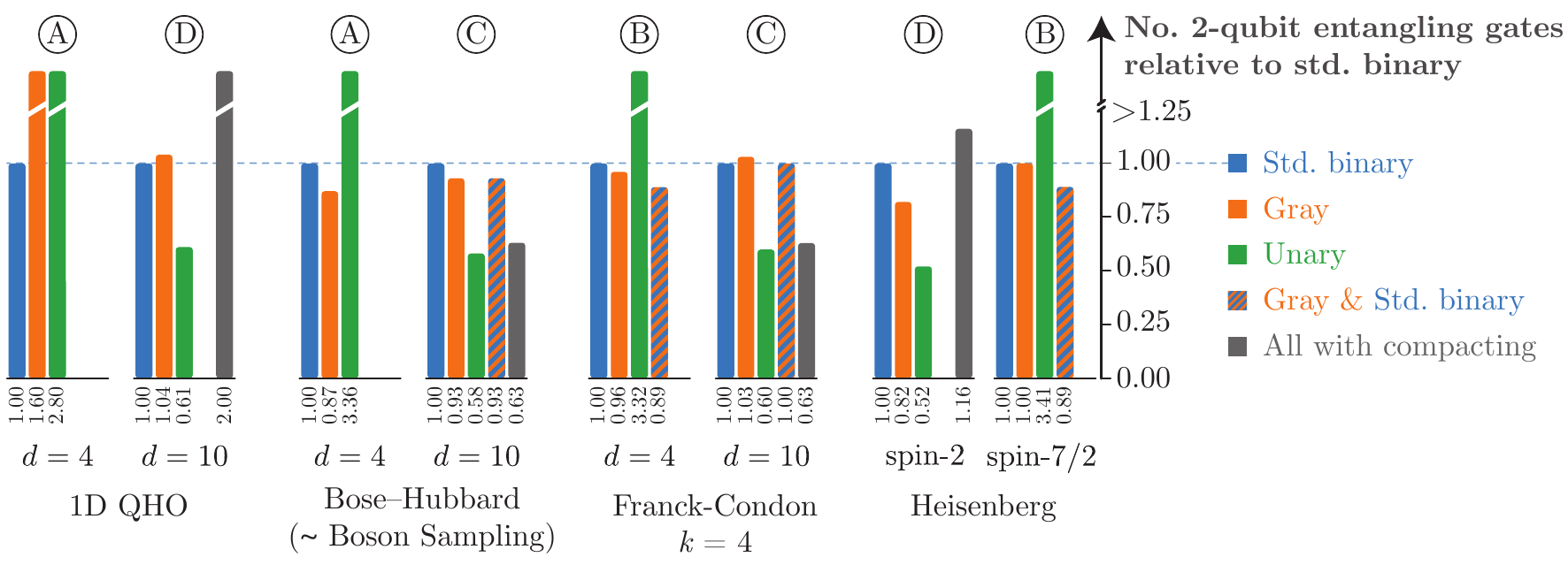}
    \caption{Using an arbitrary selection of parameters for common physics and chemistry Hamiltonians, we have plotted the comparative computational costs required for first-order Trotterization. Costs are reported in terms of number of two-qubit entangling gates, \textit{relative to} the cost of standard binary (SB). The three encodings shown here---standard binary, Gray code, and unary---are defined in the text. The five Hamiltonians are the Bose-Hubbard model, one-dimensional quantum harmonic oscillator (QHO), Franck-Condon calculation, boson sampling, and spin-$s$ Heisenberg model. The optimal encodings are sensitive both to the Hamiltonian class and the number of levels $d$ (determined by bosonic truncation or by the spin value $s$).
    In some cases, it is best to stay in a particular encoding for the duration of the simulation.
    Other times, it is worth bearing the resource cost of converting between encodings, because it saves on total operations. Still other times, the decision to save operations by converting between encodings will depend on whether available hardware is gate count limited or qubit count limited. Four Scenarios, A through D, are discussed in Section \ref{sec:physical-systems}.
}
    \label{fig:composites}
\end{figure}

Simulating quantum physics will likely be one of the first practical applications of quantum computers. In simulating the many body problem, most algorithmic progress so far has focused on systems with binary degrees of freedom, \textit{e.g.} spin-$\half$ systems \cite{Lloyd1996,childs18} or fermionic systems \cite{abrams97,bravyi02}. The latter case is relevant for simulations of chemical electronic structure \cite{cao19,mcardle18_review}, nuclear structure \cite{dumitrescu18}, and condensed matter physics \cite{wecker15}. This focus on binary degrees of freedom seems to be a natural development, partly because qubit-based quantum computation is the most widespread model used in theory, experiment, and the nascent quantum industry.


However, for a large subset of quantum physics problems, important roles are played by components that are $d$-level particles (qudits) with $d>2$, including bosonic fundamental particles \cite{fisher89}, vibrational modes \cite{wilson80}, spin-$s$ particles \cite{Levitt2008}, or electronic energy levels in molecules \cite{turro91} 
and quantum dots \cite{hong19}. 
Accordingly, several qubit-based quantum algorithms were recently developed for efficiently studying some such processes, including nuclear degrees of freedom in molecules \cite{veis2016quantum,mahesh14,teplukhin18,mcardle19,sawaya19}, the Holstein model \cite{macridin18a,macridin18b} and quantum optics \cite{sabin19,dipaolo19}.

In principle, there are combinatorially many ways to map a quantum system to a set of qubits \cite{batista04,wu02_parafermion}.
Mapping a $d$-level system to a set of qubits 
may be done by assigning an integer to each of the $d$ levels and then performing an integer-to-bit mapping.
Some consideration of $d$-level-to-qubit mappings has been published in the very recent literature, primarily for truncated bosonic degrees of freedom \cite{somma2003arXiv,veis2016quantum,macridin18a,macridin18b,mcardle19,sawaya19,sabin19}, but this is still an unexplored area of theory especially in regards to determining which encodings are optimal for which problem instances. The purpose of this work is both to provide a complete yet flexible framework for the mappings, and to analyze several encodings (both newly proposed herein and previously proposed) for a widely used set of operations and Hamiltonians. This aids in determining which mappings are more efficient for particular operators and specific hardware, including near-term intermediate-scale quantum (NISQ) devices. 


When choosing which encoding to use for a given problem, it is conceptually useful to think in terms of a \textit{hardware budget}, as shown in Figure \ref{fig:hardware_budget}.
Similar considerations have been studied for fermionic mappings \cite{steudtner18}.
For near- and intermediate-term hardware, one will often have stringent resource constraints in terms of both qubit count and gate count.
Imagine that one plans to perform Hamiltonian simulation for some $N$-particle system.
Using some set of criteria for acceptable error and other parameters, one can in principle work backwards to determine how much of a quantum resource is available for each operation.
This quantity would be different for each device. Perhaps one quantum computer would allow for more qubits but another allows for more operations, as in Figure \ref{fig:hardware_budget}.
Because different encodings yield differing resource requirements, considering multiple encodings may be essential for determining whether the available resources are sufficient.



Here we briefly summarize our results for resource comparisons of real Hamiltonian problems, in order to highlight the utility of encoding analyses and to demonstrate the ultimate practical objective of this work.
Figure \ref{fig:composites} shows the relative two-qubit operation requirements for a set of five prominent physics and chemistry problems (defined in Supplementary Section VI). All comparisons are made within a given Hamiltonian.
Our investigations revealed a somewhat rich interplay between qubit counts, operation counts, encodings, and conversions. The difficulty in \textit{a priori} predicting the optimal encoding scheme suggests that sophisticated compilation procedures, for automatically choosing and converting between multiple encodings, will play a large role in future quantum simulation efforts for $d$-level systems.




We note that the optimal encoding schemes have differing characteristics, all of which are present in Figure \ref{fig:composites}.
The results for each Hamiltonian can be categorized as one of four scenarios. \textit{Scenario A}: the optimal choice is either standard binary (SB) only or Gray only, with no benefit from converting between encodings (Bose-Hubbard $d=4$; 1D QHO $d=4$). \textit{Scenario B}: the optimal choice is to convert between SB and Gray, in order to perform different local operators in different encodings (Heisenberg $s=\frac{7}{2}$; Franck-Condon $d=4$). Scenarios A and B are notable because they require both the fewest operations and the fewest qubits, as there is no benefit to expanding into the qubit-hungry unary encoding. \textit{Scenario C}: unary-only is the optimal choice, \textit{and} saving memory by compacting the data back to Gray or SB is still cheaper than remaining in the latter encodings (Bose-Hubbard $d=10$; boson sampling $d=10$; Franck-Condon $d=10$). \textit{Scenario D}: unary is the optimal encoding, but only if the data remains in high-qubit-count unary for the duration of the calculation (1D QHO $d=10$; Heisenberg $s=2$). The optimal encoding choice is highly sensitive to both the Hamiltonian class and the truncation value $d$. This suggests the need to perform an encoding-based analysis for any new digital quantum simulation of $d$-level particles.


Throughout this work, we study four encoding types: unary (also called one-hot), standard binary, the Gray code, and a new class of encodings we name block unary, all defined in  Section \ref{sec:encoding}. After outlining how to map $d$-level operators to qubit operators for any arbitrary encoding in Section \ref{sec:mapping-operators}, we consider resource counts for most standard local operators used in bosonic and spin-$s$ Hamiltonians in Section \ref{sec:operators}. In Section \ref{sec:conversion} we present circuits for converting between encodings and enumerate their resource counts. Finally, in Section \ref{sec:physical-systems} we obtain relative resource estimates for various commonly studied Hamiltonians in theoretical physics and chemistry. Conclusions are summarized in Section \ref{sec:conclusions}. 
This work highlights how the careful choice of the encoding scheme can greatly reduce the resource requirements when simulating a system of $d$-level particles or modes.

\section{Results}

\subsection{Binary encodings}
\label{sec:encoding}





We begin by giving definitions for various integer to binary encodings. In this work, we use the term \textit{binary encoding} to refer to any code that represents an arbitrary integer as a set of ordered binary numbers, not just the familiar base-two numbering system. These encodings can be used to represent states regardless of the type of system or basis we are considering.

For a given encoding \textit{enc}, each integer $l$ has some qubit (\textit{i.e.} binary) representation, denoted 
$\mathcal{R}^{\textrm{enc}}(l)$, which is a bit string on $N_q$ bits, $x_{N_q-1} \cdots x_1 x_0$. To specify the value (0 or 1) of a particular bit $i$ in the encoding, we use the notation 
$\mathcal{R}^{\textrm{enc}}(l;i)$.

\textbf{Standard binary.} We refer to the familiar base-two numbering system as the standard binary (SB) encoding, such that an integer $l$ is represented by
\begin{equation}\label{eq:stdbin}
l \mapsto x_0 2^0 + x_1 2^1 + x_2 2^2 + ...
\end{equation}

This straight-forward mapping has been used for qubit-based quantum simulation of bosons previously \cite{veis2016quantum,mcardle19,sawaya19}. The SB mapping uses $N_q = \ceil{\log_2 d}$ qubits when the range of integers under consideration is $\{ 0,1,2,\dots,d-1\}$, where $\ceil{\cdot}$ is the ceiling function.

\textbf{Gray code.} In principle there are combinatorically many one-to-one mappings between a set of integers and a set of $N_q$-bit strings.
One mapping from classical information theory with particularly useful properties is called the Gray code or the reflected binary code \cite{roth06book}.
Its defining feature is that the Hamming distance $d_H$ between nearest-integer bit strings is always 1, formally $d_H(\mathcal{R}^{\textrm{Gray}}(l),\mathcal{R}^{\textrm{Gray}}(l+1))=1$.
The Hamming distance counts the number of mismatched bits between two bit strings. In other words, moving between adjacent integers requires only one bit flip (see Table \ref{tbl:encodingsbasic}).
As will be seen below, this encoding is especially favorable for tridiagonal operators with zero diagonals, since all non-zero elements $|l\ra\la l'|$ then have hamming distance one.
As far as we are aware, the Gray code has not been previously proposed for use in the simulation of bosons and other $d$-level systems on a qubit-based quantum computer.
This encoding inspires the possibility of having a specialized encoding for many possible matrix sparsity patterns (for example a code for which $d_H=d_H( \mathcal{R}^{\textrm{Gray}}(l) \mathcal{R}^{\textrm{Gray}}(l+1) )=2$ to be used for pentadiagonal matrices like $\hat q^2$), but we do not consider this possibility here. 
Throughout this work we refer to the SB and Gray codes as \textbf{compact encodings} because they make use of the full Hilbert space of the qubits used.


\textbf{Unary encoding.} Mappings that do not use all $2^{N_q}$ states of the Hilbert space are possible.
In the unary encoding (also called \textit{one-hot} \cite{chancellor19}, single-excitation subspace \cite{geller19}, or direct mapping \cite{mcardle19}), 
the number of qubits required is $N_q = d$.
Therefore, only an exponentially small subspace of the qubits' Hilbert space is used.
The relationship between the unencoded and encoded ordered sets is

\begin{equation}
\texttt{\{0, 1, 2, 3, ...\}} \mapsto \texttt{\{...000001, ...000010, ...000100, ...001000, ...\}}.
\end{equation}

Previous proposals for bosonic simulation on a universal quantum computer have used this encoding \cite{somma05,geller19,mcardle19,sawaya19,chancellor19} under different names.
The unary encoding makes less efficient use of quantum memory, but it will become clear below that it usually allows for fewer quantum operations.

\textbf{Block unary encoding.} One can interpolate between the two extremes, using less than the full Hilbert space but more of the space than the unary code uses.
In some limited instances this allows one to make tunable trade-offs between required qubit counts and required operation counts, which may be especially useful for mapping physical problems to the specific hardware budget of a particular near-term intermediate scale (NISQ) device. 
In this work, we introduce a class of such encodings that we call block unary.

The block unary code is parameterized by choosing an arbitrary compact encoding (\textit{e.g.} SB or Gray code) and a local parameter $g$ that determines the size of each block.
It can be viewed as an extension of the unary code, where each digit (block) ranges from 0 to $g$.
Within each block, the local encoding is used to represent the local digit. The number of qubits required is $N_q = \ceil{\frac{d}{g}} \ceil{\log_2 (g+1)}$. 
Examples of the block unary encoding are given in Table \ref{tbl:blockunary}.
We use the notation BU$^{\textrm{enc}}_g$ to define block unary with a particular pair of parameters.
For a transition within a particular block, the number of operations is similar to a compact code with $d=g+1$.
For elements that move between different blocks, the transition will be conditional on all qubits in both blocks.


\textbf{Bitmask subset.} Because some encodings do not make use of the full Hilbert space, it will be useful to define the subset of bits that is necessary to determine each integer $l$.
For a given encoding we call this subset the \textit{bitmask subset}, denoted $C(l)$ where $C_l \subseteq \{0,1,2,...,N_q-1\}$.
The bitmask subset for the SB and Gray encodings is always $C^{\textrm{SB}}(l) = C^{\textrm{Gray}}(l) = \{0,1,2,...,N_q-1\}$, since all $N_q$ bits must be known to determine the integer value.
In the unary encoding, the bitmask subset is simply $C^{\textrm{Unary}}(l)=\{l\}$, because if one knows that bit $l$ is set to 1, then one knows the other bits are 0.
In the block unary code, the bitmask subset simply contains the bits that represent the current block.
For example, for the Block Unary Gray code with $g=3$ (see Table \ref{tbl:blockunary}), $C^{\textrm{BU}[g=3]}(2) = \{0,1\}$ and $C^{\textrm{BU}[g=3]}(3) = \{2,3\}$.



\begin{table}[]
\begin{tabular}{|l|c|c|c|}
Decimal & SB & Gray & Unary \\
0  & \texttt{0000}  & \texttt{0000} & \texttt{000000000001}  \\
1  & \texttt{0001}  & \texttt{0001} & \texttt{000000000010}  \\
2  & \texttt{0010}  & \texttt{0011} & \texttt{000000000100}  \\
3  & \texttt{0011}  & \texttt{0010} & \texttt{000000001000}  \\
4  & \texttt{0100}  & \texttt{0110} & \texttt{000000010000}  \\
5  & \texttt{0101}  & \texttt{0111} & \texttt{000000100000}  \\
6  & \texttt{0110}  & \texttt{0101} & \texttt{000001000000}  \\
7  & \texttt{0111}  & \texttt{0100} & \texttt{000010000000}  \\
8  & \texttt{1000}  & \texttt{1100} & \texttt{000100000000}  \\
9  & \texttt{1001}  & \texttt{1101} & \texttt{001000000000}  \\
10 & \texttt{1010} & \texttt{1111} & \texttt{010000000000}  \\
11 & \texttt{1011} & \texttt{1110} & \texttt{100000000000}  
\end{tabular}
\caption{The standard binary (SB), Gray code, and unary encodings. We refer to the SB and Gray codes as \textit{compact} encodings.}
\label{tbl:encodingsbasic}
\end{table}

\begin{table}[]
\begin{tabular}{|l|c|c|c|c|}
Decimal &     BU$_{g=3}^\textrm{SB}$  &    BU$_{g=3}^\textrm{Gray}$  &   BU$_{g=5}^\textrm{Gray}$ &     BU$_{g=7}^\textrm{Gray}$    \\
0  & \texttt{00 00 00 01}   & \texttt{00 00 00 01}  & \texttt{000 000 001} & \texttt{000 001}  \\
1  & \texttt{00 00 00 10}  & \texttt{00 00 00 11} & \texttt{000 000 011} & \texttt{000 011}  \\
2  & \texttt{00 00 00 11}  & \texttt{00 00 00 10} & \texttt{000 000 010} & \texttt{000 010}  \\
3  & \texttt{00 00 01 00}  & \texttt{00 00 01 00} & \texttt{000 000 110} & \texttt{000 110}  \\
4  & \texttt{00 00 10 00}  & \texttt{00 00 11 00} & \texttt{000 000 111} & \texttt{000 111}  \\
5  & \texttt{00 00 11 00}  & \texttt{00 00 10 00} & \texttt{000 001 000} & \texttt{000 101}  \\
6  & \texttt{00 01 00 00}  & \texttt{00 01 00 00} & \texttt{000 011 000} & \texttt{000 100}  \\
7  & \texttt{00 10 00 00}  & \texttt{00 11 00 00} & \texttt{000 010 000} & \texttt{001 000}  \\
8  & \texttt{00 11 00 00}  & \texttt{00 10 00 00} & \texttt{000 110 000} & \texttt{011 000}  \\
9  & \texttt{01 00 00 00}  & \texttt{01 00 00 00} & \texttt{000 111 000} & \texttt{010 000}  \\
10 & \texttt{10 00 00 00}  & \texttt{11 00 00 00} & \texttt{001 000 000} & \texttt{110 000}  \\
11 & \texttt{11 00 00 00}  & \texttt{10 00 00 00} & \texttt{011 000 000} & \texttt{111 000}  
\end{tabular}
\caption{Examples of the block unary SB and block unary Gray encodings.}
\label{tbl:blockunary}
\end{table}


\subsection{Mapping $d$-level Matrix Operators to Qubits}
\label{sec:mapping-operators}


Any operator for a $d$-level system can be written as
\begin{equation}\label{eq:genOp}
\hat A = \sum_{l,l'=0}^{d-1} a_{l,l'} | l \ra\la l' |,
\end{equation}
where $l$ and $l'$ are integers labelling pairwise orthonormal quantum states. In this work we conceptualize the mappings primarily in terms of \textit{Fock-type} encodings (or alternatively \textit{second quantization} encodings) where each $|l\ra$ represents one level in the $d$-level system. However, the mapping procedure is identical to the one used for first quantization operators \cite{macridin18a,macridin18b} that we briefly discuss. When dealing with bosonic degrees of freedom, one must choose an arbitrary level $d$ at which to truncate, since in principle a bosonic mode may have an unbounded particle number. Choosing this cutoff such that truncation error is below a given threshold is an essential step that has been previously studied \cite{lee14_truncated,woods15_bosonic,sawaya19}, though it is beyond the scope of the current work.

In performing a mapping of any $d$-by-$d$ matrix operator to a sum of Pauli strings, the following approach may be used.
For each term in the sum, one first assigns an integer to each level and then uses an arbitrary binary encoding $\mathcal{R}$ to encode each integer:

\begin{equation}
|l\ra \mapsto |\mathcal{R}(l)\ra = |\mathcal{R}(l;N_q-1)\ra\cdots|\mathcal{R}(l;1)\ra|\mathcal{R}(l;0)\ra
\mapsto |x_{N_q-1}\ra \cdots |x_0\ra, \; x_i \in \{0,1\}.
\end{equation}
For codes using less than the full Hilbert space of the qubits (\textit{e.g.} unary and block unary), some qubits can be safely ignored for a given element $\llp$. This is because operations on these excluded qubits will not affect the manifold on which the problem is encoded. Therefore, for each element $|l \ra\la l'|$ a mapping needs to consider only the bitmask subsets of the two integers, ignoring other bits. The operator on the qubit space is then

\begin{equation}
| l \ra\la l' | \mapsto \bigotimes_{i \in C(l) \cup C(l')} | x_i \ra\la x'_i |_i,
\end{equation}

where $C(l) \cup C(l')$ is the union of the bitmask subsets of the two integers and the subscripts $i$ denote qubit number. 
One then converts each qubit-local term $|x_j \ra\la x_j'|$ to qubit operators using the following four expressions:

\begin{equation}\label{eq:op01}
|0\ra\la 1| = \half ( \sx + i\sy ) \equiv \hat \sigma^-, 
\end{equation}

\begin{equation}\label{eq:op10}
|1\ra\la 0| = \half ( \sx - i\sy ) \equiv \hat \sigma^+, 
\end{equation}

\begin{equation}\label{eq:op00}
|0\kb 0| = \half (I + \sz),
\end{equation}

\begin{equation}\label{eq:op11}
|1\kb 1| = \half (I - \sz).
\end{equation}

For a single term in equation \eqref{eq:genOp}, the result is a sum of Pauli strings,

\begin{equation} \label{eq:sum-of-pauli}
\hat A \mapsto \sum_k^P c_{k} \bigotimes_j^{N_q} \hat \sigma_{kj},
\end{equation}
where $P$ is the number of Pauli strings in the sum, $c_{k}$ is a coefficient for each Pauli term, and every operator is either a Pauli matrix or the identity: $\hat \sigma_{kj} \in \;\{\sx,\;\sy,\;\sz\}\cup \{I\}$. Note that in this work the set of Pauli matrices is defined to exclude the identity.



\subsection{Significance of the Hamming distance}

It is useful to analyze encoding efficiency based on the \textit{Hamming distance} between $\mathcal{R}(l)$ and $\mathcal{R}(l')$.
The Hamming distance, which we denote $d_H^{\mathcal R}(l,l') \equiv d_H(\mathcal R(l),\mathcal R(l'))$, is defined as the number of unequal bits between two bit strings of equal length.
The important observation is that, for a given element $a_{l,l'} | l \ra\la l' |$ in equation \eqref{eq:genOp}, the average length of the Pauli strings increase as the Hamming distance increases, where length is defined as the number of Pauli operators (excluding identity) in the term.
In this subsection, for simplicity we at first assume that the bitmask subset is $C(l)=\{0,1,...,N_q-1\}$, implying that we are using a compact code such as Gray or SB. But we note that these Hamming distance considerations are relevant to all encodings.
In the case of a non-compact encoding, one would consider only the union of the bitmask subsets.
To clarify this result, consider the following. For an arbitrary element $| l \ra\la l' |$ written in binary form $| \mathcal R(l) \ra\la \mathcal R(l') |$, one performs the following mapping:

\begin{equation} \label{eq:pauliexpand}
|x_0 \kb x_0'|\otimes\cdots\otimes|x_N \kb x_N'| \mapsto \frac{1}{2^N} (\hat a_0+\hat b_0) (\hat a_1+\hat b_1) \cdots (\hat a_N+\hat b_N)
\end{equation}

where subscripts denote qubit number, $\hat a \in \{\sx, I\}$, and $\hat b \in \{\pm i\sy,\pm \sz\}$ (according to equations \eqref{eq:op01} through \eqref{eq:op11}). Expanding the RHS of equation \eqref{eq:pauliexpand} leads to an equation of form \eqref{eq:sum-of-pauli}.
For any mismatched qubit $j$, \textit{i.e.} any qubit for which $x_j \neq x_j'$, the clause $(\hat a_j + \hat b_j)$ contains two Pauli operators. For matched qubits ($x_j = x_j'$), the clause $(\hat a_j + \hat b_j)$ instead has one identity and one Pauli operator. Hence more \textit{matched} bits lead to more identity operators in expression \ref{eq:pauliexpand}, leading to fewer Pauli operators in the final sum of Pauli strings. It follows that the number of non-identity operators can be reduced by having a smaller Hamming distance. This is relevant because Hamiltonians with more Pauli operators require more quantum operations to implement.


Consider the illustrative example of mapping the Hermitian term $|3\ra\la4| + |4\ra\la3|$ to a set of qubits.
In the SB encoding, equations \eqref{eq:op01}--\eqref{eq:op11} yield the following Pauli string representation:
\begin{equation}\label{eq:sb_example_34}
\begin{split}
&  |3\ra\la4| + |4\ra\la3|  \xmapsto{\textrm{Std. Binary}} |011 \ra\la 100| + |100 \ra\la 011| \\
&  = \frac{1}{4} ( \sx^{(2)}\sx^{(1)}\sx^{(0)} + \sy^{(2)}\sy^{(1)}\sx^{(0)} + \sy^{(2)}\sx^{(1)}\sy^{(0)} - \sx^{(2)}\sy^{(1)}\sy^{(0)} ).
\end{split}
\end{equation} 

Using the Gray code, the Pauli string instead takes the form
\begin{equation}\label{eq:gray_example_34}
\begin{split}
  &|3\ra\la4| + |4\ra\la3|  \xmapsto{\textrm{Gray}} |010 \ra\la 110| + |110 \ra\la 010| \\
  &= \frac{1}{4} ( - \sx^{(2)} \sz^{(1)} \sz^{(0)} + \sx^{(2)} \sz^{(0)} - \sx^{(2)} \sz^{(1)} + \sx^{(2)} ).
\end{split}
\end{equation}

The Hamming distance between $\mathcal R(3)$ and $\mathcal R(4)$ is $d_H^{\textrm{SB}}=3$ in the former case and $d_H^{\textrm{Gray}}=1$ in the latter. The result is that the Gray code has fewer Pauli operators per Pauli string, meaning that it can be implemented with fewer operations.








\begin{figure}[ht]
\centering
\includegraphics[width=0.85\textwidth]{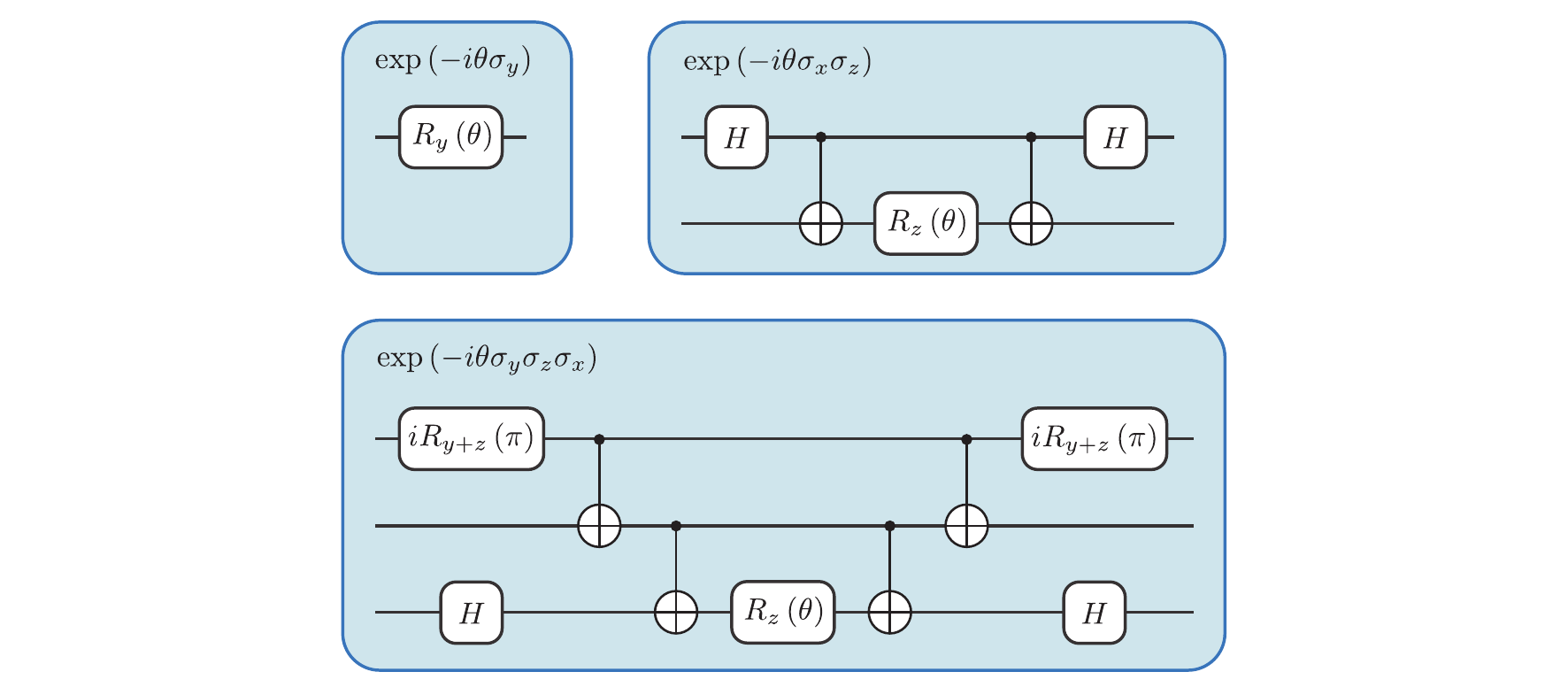}
\caption{Canonical quantum circuits used to exponentiate Pauli strings on a universal quantum computer. One needs $2(p-1)$ two-qubit gates for such an operation, where $p$ is the number of Pauli operators in the term.  
When a product of many exponentials is used, as in the Suzuki-Trotter procedure, there tends to be significant gate cancellation.
}
\label{fig:cnot_stairs}
\end{figure}

\subsection{Avoiding superfluous terms in non-compact codes}

When implementing local products of operators using non-compact codes (unary or BU), one should multiply operators in the matrix representation before performing the encoding to qubits. If one instead first maps local operators to qubit operators, and then multiplies the operators, superfluous terms may result. For example, when implementing an arbitrary squared operator $\hat A^2$, one should begin with the matrix representation of $\hat A^2$ instead of squaring the qubit representation of $\hat A$. To see this, consider the unary encoding of the square of a 3-level matrix operator $\hat n_{d=3} = \textrm{diag} [0,1,2] $ $\mapsto \frac{3}{2} I - \half \sz^{(1)} - \sz^{(2)}$. If one begins with $\hat n_{d=3}^2 = \textrm{diag} [0,1,4]$, the encoded operator is 

\begin{equation}\label{eq:n3sq_mult_first}
\hat n_{d=3}^2 = \textrm{diag} [0,1,4]  \xmapsto{\textrm{Unary}}  \frac{5}{2} I - \half \sz^{(1)} - 2 \sz^{(2)}.
\end{equation}

If one instead squares the already-encoded Pauli operator for $\hat n_{d=3}$, this yields
\begin{equation}\label{eq:n3sq_map_first}
(\frac{3}{2} I - \half \sz^{(1)} - \sz^{(2)})^2 = \frac{7}{2} I - \frac{3}{2} \sz^{(1)} + \sz^{(1)}\sz^{(2)} - 3\sz^{(2)}.
\end{equation}

Superscripts denote qubit number. Pauli operators \eqref{eq:n3sq_mult_first} and \eqref{eq:n3sq_map_first} behave identically on the subspace of the unary encoding, though operator \eqref{eq:n3sq_mult_first} is less costly to implement. One might attempt to eliminate superfluous terms after the mapping is complete, but this is likely a hard problem. In principle it may require combinatorial effort to determine which combinations of operators leave the encoding space unaffected. Hence the most prudent strategy is to always perform as much multiplication as possible in the matrix representation. These considerations are irrelevant when using one of the compact encodings.

\subsection{Trotterization and gate count upper bounds}

Hamiltonian simulation often consists of implementing the unitary operator
\begin{equation}\label{eq:propagator}
\hat U(t) = \exp(-i \hat H t)
\end{equation}
for some user-defined time-independent Hamiltonian $\hat H$, where $t$ is the evolution time and we have set $\hbar=1$. Any Hamiltonian can be expressed as a sum of local Pauli strings such that 
\begin{equation}
\hat H = \sum_k c_k \bigotimes_j \hat \sigma_{kj} = \sum_k \hat h_k,
\end{equation}
which takes the same form as equation \eqref{eq:sum-of-pauli} with the Pauli strings and their coefficients compacted into terms $\{\hat h_k\}$. In practice, Hamiltonian simulation can be performed using a Suzuki-Trotter decomposition
\begin{equation}\label{eq:trotter}
\hat U(t) = \exp\left(-i t \sum_k \hat h_k\right)
     \approx \left(\prod_k \exp(-i \hat h_k t/\eta) \right)^\eta
     = \tilde U(t),
\end{equation}
where the expression is exact in the limit of large $\eta$ or small $t$ \cite{suzuki76,Lloyd1996}. The numerical studies of this work consider the encoding-dependent resource counts for equation \eqref{eq:trotter}, for a subset of prominent physics problems. We focus on determining the resources required for simulating a single Trotter step.

%


There are several variations and extensions to the Hamiltonian simulation approach of equation \eqref{eq:trotter}, including higher-order Suzuki-Trotter methods \cite{suzuki76}, the Taylor  series algorithm \cite{Berry2015}, quantum signal processing \cite{qsp2017}, and schemes based on randomization \cite{Childs2019,Campbell2019}. Notably for the current work, recent results suggest that simple first-order Trotterization will have lower error for near- and medium-term hardware \cite{Childs2018,childs19_theoryoftrotter}, even if the other methods are asymptotically more efficient.
Since $\hat h_k$ takes a different form depending on the chosen encoding, the resource counts as well as the error $|\hat U(t)-\tilde U(t)|$ will be different.
We leave the study of numerical error for future work, as the goal of the current work is to introduce these mappings and to understand some trends in their resource requirements.

Each term $\exp(-i t \hat h_k)=\exp(-i t c_k \bigotimes_j \hat \sigma_{kj})$ may be implemented using the well-known CNOT staircase quantum circuit shown in Figure \ref{fig:cnot_stairs}.
If a qubit $j$ is acted on by $\sx$ or $\sy$, additional single-qubit gates are placed on qubit $j$ as shown in the figure.
$H \equiv i R_{x+z}(\pi)$ denotes the Hadamard gate that changes between the Z- and X-basis, and $i R_{x+y}(\pi)$ converts between the Z- and Y-basis.
To exponentiate a single Pauli string, the number of CNOT (CX) gates required is 
\begin{equation}\label{eq:ncxub_sgl_pauli}
n_{\textrm{cx}}(p) = 2(p-1),  
\end{equation}
where $p$ is the number of Pauli operators ($\{\sx,\sy,\sz\}$, excluding $I$) in the term to be exponentiated.

\begin{figure}[h]
    \centering
    \includegraphics[width=0.5\textwidth]{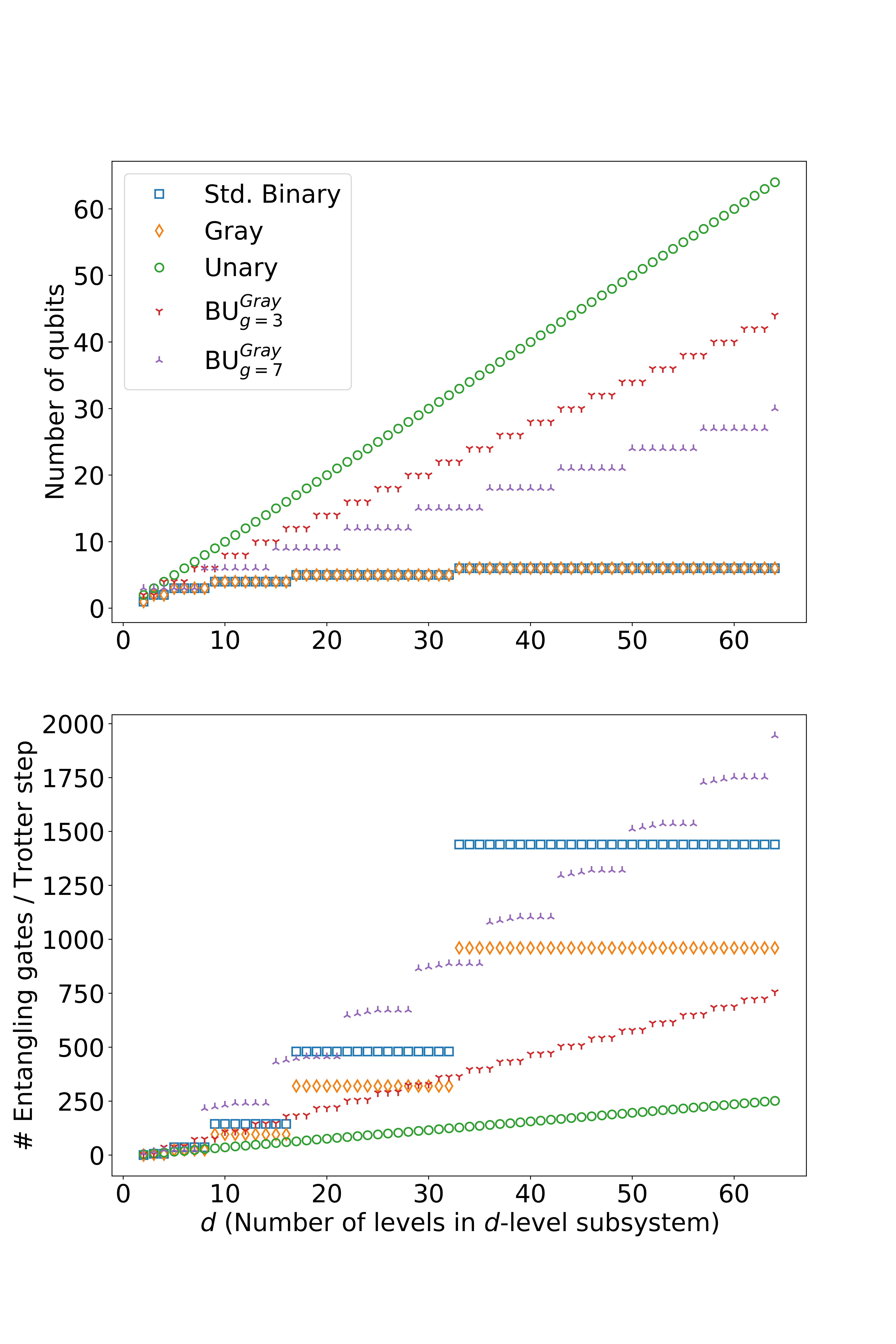}
    \caption{Numerical upper bounds for resource counts of implementing one Suzuki-Trotter step of a $d$-by-$d$ real Hermitian matrix operator $\hat B$, where $\hat B$ is tridiagonal with zeros on the diagonal. Top: Qubit counts for mappings considered in this work. BU$^{Gray}_{g}$ stands for block unary where $g$ is the size of the block. Asymptotically, the number of qubits scales logarithmically for the SB and Gray encodings, and linearly for the unary and block unary encodings. Bottom: Upper bounds of CNOT operation counts for implementing one Suzuki-Trotter step of $\hat B$. This is the sparsity pattern of canonical bosonic position and momentum operators as well as the $S_x$ spin operators in spin-$s$ systems. Upper bounds were calculated by mapping the full operator to a sum of weighted Pauli strings, combining terms, and then using equation \eqref{eq:ncxub_sgl_pauli}. Notably,  encodings with higher qubit counts tend to have lower upper bounds for gate counts, and vice-versa.}
    \label{fig:count_qub_op_ub}
\end{figure}
%

It is instructive to calculate upper bounds for entangling gate counts. Consider a simple Hermitian term $\alpha\llp$ + $\alpha^*\lpl$ and a single diagonal element $\lkbl$. 
Here, $d_H$ denotes $d_H(\mathcal R(l),\mathcal R(l'))$, which will depend on the chosen encoding, and $d_H=0$ in the case of a diagonal element. 
We define $K$ as $|C^{\textrm{enc}}(l) \cup C^{\textrm{enc}}(l')|$, the number of qubits in the relevant bitmask subsets. As we are considering products of two-term sums (equations \eqref{eq:op01}--\eqref{eq:op11}), the distribution of Pauli strings can be analyzed in terms of binomial coefficients. In the Supplementary Section I we show that 
\begin{equation}\label{eq:2lev_cx_ub}
\begin{split}
n_{\textrm{cx,UB}}[d_H(l,l'),K] = \half \sum_{p=2}^{K} 2^{d_H} {K-d_H \choose p-d_H} (2p-2),
\end{split}
\end{equation}
where \textit{UB} denotes upper bound and the $\half$ factor is not present for diagonal terms. Some resulting upper bounds for particular Hamming distances are

\begin{equation}\label{eq:ncxub_binom_gen}
\begin{split}
n_{\textrm{cx,UB}}[d_H=0,K] = (2^K K-2(2^K) - 2)  \\
n_{\textrm{cx,UB}}[d_H=1,K] = \half (2^K K-2^K)   \\
n_{\textrm{cx,UB}}[d_H=2,K] = \half (2^K K) . \\
\end{split}
\end{equation}

Again, the above expressions are for a single Hermitian element pair or a single diagonal term. In practice, because substantial gate cancellation is possible once the quantum circuit has been compiled, these upper bounds are not always directly applicable when choosing an encoding. However, the above expressions may find direct utility in limiting cases, and they demonstrate the basic relationship between Hamming distance, size of the bitmask union, and gate counts.
Below, we study a common sparsity pattern, a tridiagonal real matrix operator $\hat B$ with zeros on the diagonal, \textit{i.e.} with matrix structure $\la i | \hat B | j \ra= \sum_k B_k ( \delta_{i,k}\delta_{j,k+1} + \delta_{i,k+1}\delta_{j,k}$). 
This is the sparsity pattern of several commonly used $d$-level operators, such as the bosonic position operator $\hat q$.

\begin{table}[]
\begin{tabular}{lccc}
                     & \hspace{1mm} $\alpha\llp + \alpha^*\lpl$ \hspace{1mm}  & $\hat B$ & Dense \\ \hline
Unary               &   $O(1)$    & $O(d)$  &  $O(d^2)$ \\
Block unary         &   $O(g \log g)$ & $O(d g \log g)$ & \hspace{2mm}$O(d^2 g \log g)$\hspace{2mm} \\
Compact (SB or Gray)&   $O(d \log d)$    & $O(d^2 \log d)$  & \hspace{2mm}$O(d^2 \log d)$\hspace{2mm}      \\ \hline
\end{tabular}
\caption{Asymptotic upper bound complexity of entangling gate counts, for Trotterizing a matrix exponential of a $d$-level particle. The second column refers to one Hermitian matrix element pair; the third column to a tridiagonal matrix operator $\hat B$ with zeros on the diagonal; the last column to a dense matrix operator. These asymptotic complexities are useful primarily for considering general trends---for smaller values of $d$, it is best to numerically test all encodings to determine which requires the fewest operations.}
\label{tbl:cnot-ub}
\end{table}


In Table \ref{tbl:cnot-ub} we show analytical upper bounds for three different levels of sparsity, derived in Supplementary Section I. We consider a single Hermitian pair, the $\hat B$ operator with $O(d)$ nonzero entries, and a dense matrix operator with $O(d^2)$ nonzero entries. In Supplementary Section I we show that the upper bound of entangling gate counts for an arbitrary operator on $K$ qubits is $O(K4^K)$. Because $K = \ceil{\log_2 d}$ in the compact codes, this means that the compact codes will never require more than $O(d^2 \log d)$ entangling gates.

An important consequence is that, as the matrix density increases, the comparative advantage of the unary encoding decreases. The compact codes' upper bound both for a $\hat B$ and for a fully dense operator are both $O(d^2 \log d)$, since this is the maximum upper bound. The unary encoding's upper bound of $O(d^2)$ for fully dense operator is only slightly lower. And because actual gate counts for various matrix instances will be less than these upper bounds, it appears possible that compact codes might often be superior for dense matrices in both qubit count and gate count. However, because the most commonly used quantum operators tend to have $O(d)$ density, it is likely that unary will most often be superior in gate counts, at least for Hamiltonian simulation. Many exceptions to these trends are shown in Section \ref{sec:operators}.

As a more concrete demonstration of typical operator scaling, we calculated numerical upper bounds for $\hat B$ with increasing $d$. Qubit counts and upper bounds for $\hat B$ are shown in Figure \ref{fig:count_qub_op_ub}, as a function of $d$ for different encodings. We first encoded the entire operator $\hat B$ into a sum of Pauli strings before collecting and cancelling terms, leading to some favorable cancellations. 
Then we applied equation \eqref{eq:ncxub_sgl_pauli}.
There is roughly an inverted relationship between the qubit counts and the operation counts, because sparser encodings like the unary and block unary have smaller bitmask subsets but require more total qubits.

The differing gate count upper bounds between the SB and gray encodings (Figure \ref{fig:count_qub_op_ub}) are explained by Hamming distances.
Because all non-zero terms in $\hat B$ have unity Hamming distance, upper bounds for the Gray code are substantially lower.
The other notable trend is that the unary code has lower upper bounds, asymptotically, than the other codes. This can be explained using equation \eqref{eq:ncxub_binom_gen} by noting that $K=2$ for any matrix element, while $K=\ceil{\log_2 d}$ for Gray and SB. In other words, $K$ stays constant in the unary encoding, whereas in the compact codes $K$ increases with $d$. Upper bounds for BU$^{Gray}_{g=3}$ are between the compact codes and the unary code, as this encoding has an intermediate value of $K$.
Below we will see that, though these trends generally persist, they are less pronounced and less predictable after cancelling of Pauli terms and circuit optimization.

\subsection{Diagonal Binary-Decomposable Operators }\label{sec:dbd}

\begin{figure}
    \centering
    \includegraphics[width=0.6\textwidth]{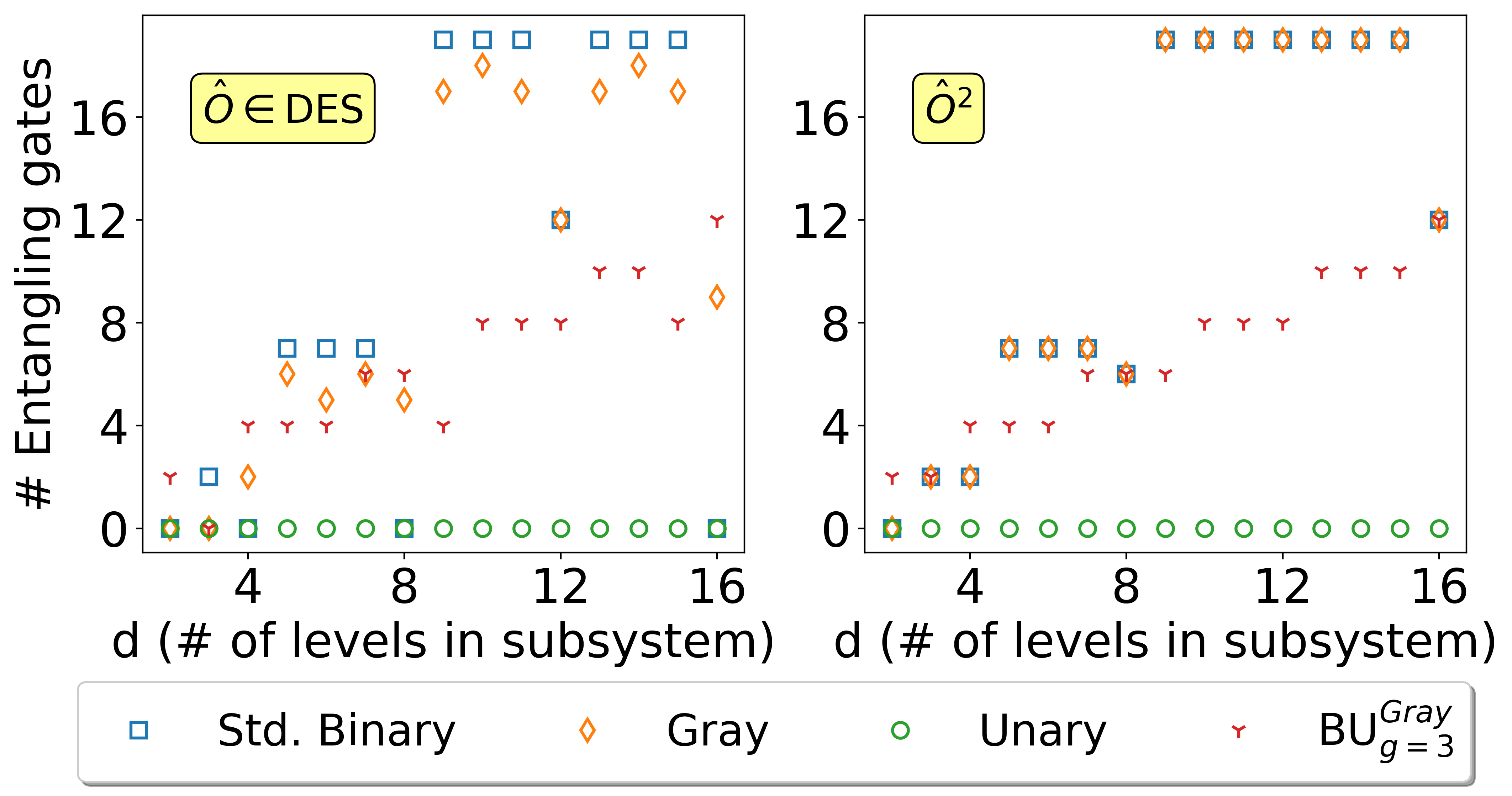}
    \caption{Entangling gates for a single Suzuki-Trotter step of an arbitrary diagonal evenly-spaces (DES) operator $\hat O$ (left), and its square (right). Gate counts are for optimized quantum circuits. DES operators are a subset of the diagonal binary-decomposable (DBD) operator class. Because it is diagonal, the unary code always requires only single-qubit operations. When $\log_2 d$ is an integer, the SB code requires no entangling gates and just $\log_2 d$ single-qubit operations, making it the most efficient encoding (in terms of both qubits and operations). DBD operators are a common operator class, encompassing \textit{e.g.} the bosonic number operator $\hat n$ and the spin-$s$ operator $\hat S_z$.}
    \label{fig:dbd}
\end{figure}


An important class of operators to consider is those which we call diagonal binary-decomposable (DBD). We define DBD operators as being diagonal matrix operators for which the diagonal entries of the operator ($\hat O$) may be expressed as 


\begin{equation}\label{eq:dbd}
\hat O_{l,l} = \sum_{i=1}^{\ceil{\log_2d}} k_i R^{\textrm{SB}}(l;i) 
\end{equation}

where $R^{\textrm{SB}}(l;i) \in \{0,1\}$. A common subclass of DBD is the set of diagonal operators containing evenly-spaced entries. We call these diagonal evenly-spaced (DES) operators. An example is the bosonic number operator

\begin{equation}\label{eq:n_explicit}
\hat n = \textrm{diag} [0, 1, 2, 3, \cdots]
\end{equation}

and any linear combination $a \hat n + b I$ where $a$ and $b$ are constants. If $\log_2 d$ is an integer, then the Pauli operator is simply the base-two numbering system with $k_i=2^i$, 

\begin{equation}
\hat n = 2^0 \sz^{(0)} + 2^1 \sz^{(1)} + 2^2 \sz^{(2)} + \cdots
\end{equation}

The DBD class of operators is notable because, when $\log_2 d$ is an integer, \textit{exactly} implementing $\exp(-i \theta \hat n)$ requires only $\log_2 d$ single-qubit rotations and no entangling gates. 

An operator for which $\log_2 d$ is a non-integer will not lead to this favorable only-single-qubit decomposition. For example, the $\hat S_z$ operator for a spin-$s$ system is DBD, but the advantage appears for the SB mapping only when $d=2s+1$ is a power of 2, namely $s \in \{3/2, 7/2, \dots\}$. However, for other operators one may simply increase $d$ without changing the simulation result. For example, if one is required to exponentiate a truncated bosonic operator $\hat n$ with at least $d=11$, it is most efficient simply to implement the SB encodings of $\hat n$ with $d=16$ instead. A simple example illustrates this point. The standard operator for the bosonic number operator with truncation $d$=3 is
\begin{equation}\label{eq:dbd_sb_n3}
\hat n_{d=3} = \textrm{diag} [0,1,2] \xmapsto{\textrm{Std. Binary}} \frac{3}{4}I + \frac{1}{4}\sz^{(0)} - \frac{3}{4}\sz^{(0)}\sz^{(1)} - \frac{1}{4}\sz^{(1)}
\end{equation}
while that for $d$=4 is
\begin{equation}\label{eq:dbd_sb_n4}
\hat n_{d=4} = \textrm{diag} [0,1,2,3] \xmapsto{\textrm{Std. Binary}} \frac{3}{2}I + \frac{1}{2}\sz^{(0)} - \sz^{(1)}
\end{equation}

The latter operator ($d=4$) is composed only of single-qubit operators but the former ($d=3$) is not. Operations counts for CNOT gates are shown in Figure \ref{fig:dbd}, where it is clear that the SB mapping is superior when $d$ is a power of 2. The right panel gives gate counts for operators such as $\hat n^2$, where it is again advangatgeous for $d$ to be a power of 2, though entangling gates are still required. As is also clear from the right panel, the square of a DES or DBD operator is in general not DBD.



\subsection{Local operators}
\label{sec:operators}


%
\begin{figure}
    \centering
    \includegraphics[width=0.7\textwidth]{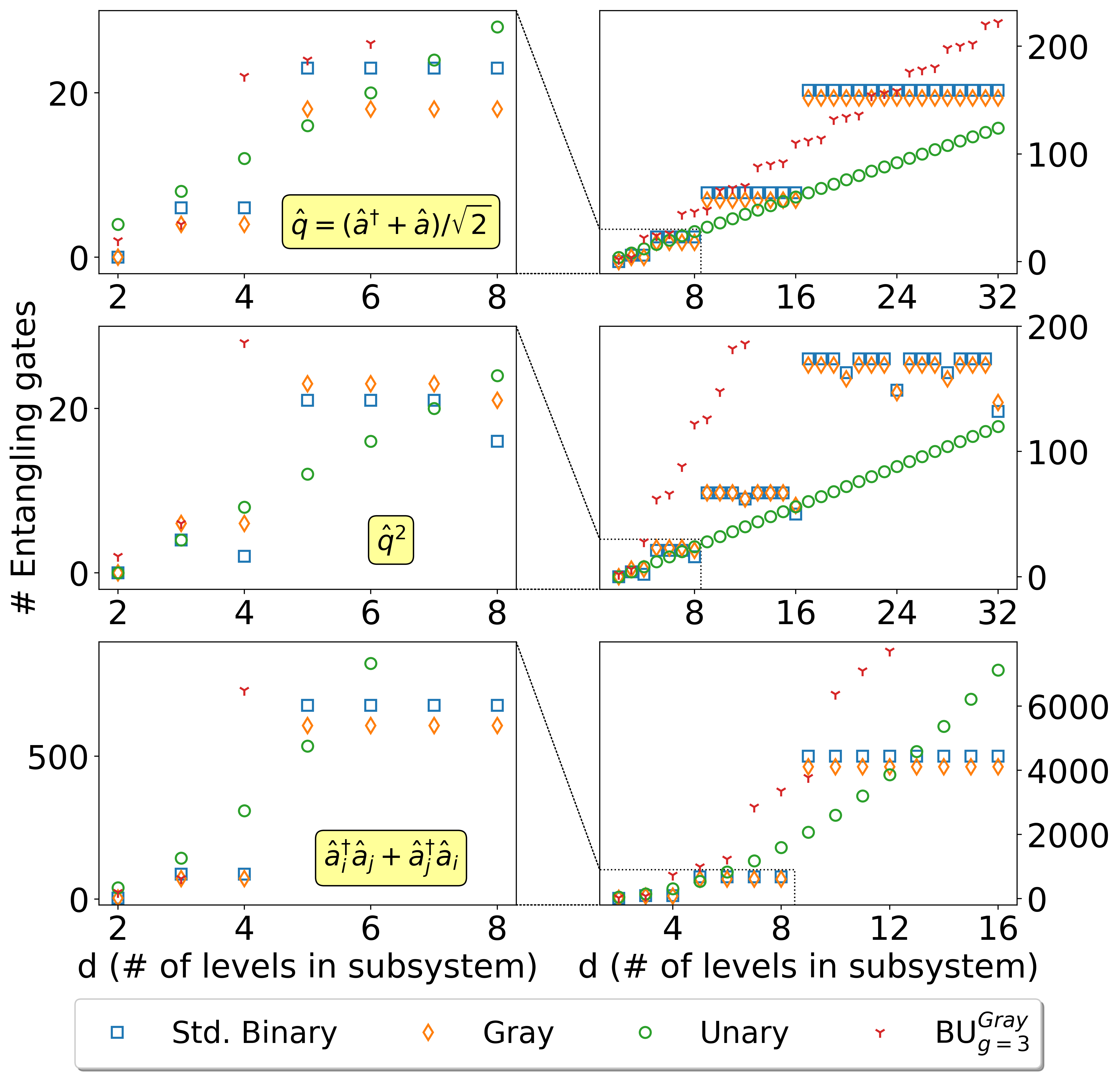}
    \caption{CNOT gate counts for optimized circuits of the bosonic position operator (first row), position operator squared (second row), and two-site bosonic interaction operator (third row). Plots on the left correspond to enlargements of the dotted boxes in the plots on the right. There are several notable trends and anomalies: (a) Though unary usually requires the fewest operations as $d$ increases, there are several operators for which the Gray or SB code is more efficient than the unary in \textit{both} qubit and operation count. This occurs most pronouncedly at values such as $d=4$, 7, and 8. (b) The Gray code is usually more efficient than SB even after circuit optimization, especially for operators composed of tridiagonal operators (top and bottom rows). (c) The Gray code's advantage is either less pronounced or disappears for $\hat q^2$, because $\hat q^2$ is a pentadiagonal operator, for which the unity Hamming distance of the Gray code is less useful. (d) The reduction in operation count for $\hat q^2$ at values such as $d=8$, 24, or 32 occur because the diagonal of $\hat q^2$ is DBD. 
    }
    \label{fig:loc_d8}
\end{figure}
\begin{figure}
    \centering
    \includegraphics[width=0.7\textwidth]{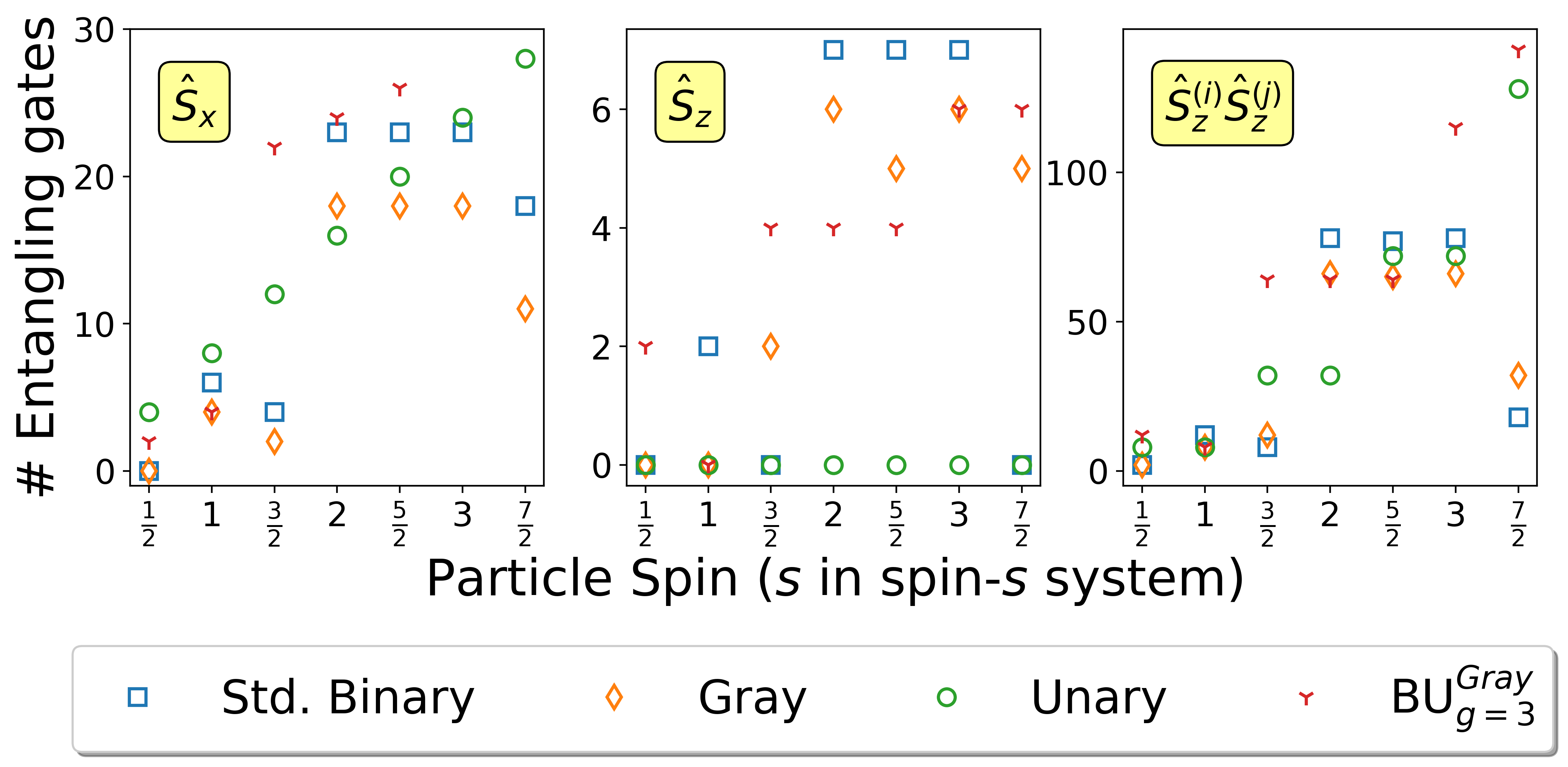}
    \caption{ CNOT gate counts of optimized Suzuki-Trotter circuits, for approximating the exponentials of the spin-$s$ operators shown. There is no clean overall trend for $\Sx$ and $\Sint$ (except that Gray tends to out-perform SB), highlighting the need to study encodings thoroughly for each new use case. Notably, because $\hat S_z$ is diagonal binary-decomposable, for values of $s= \frac{3}{2} , \frac{7}{2} $ (which are 4- and 8-level system respectively) the SB code requires both fewest operations and fewest qubits. 
    }
    \label{fig:locspin}
\end{figure}

Though local $d$-level operators can in principle contain arbitrary terms and even be entirely dense (\textit{i.e.} a molecule's electronic energy levels with non-zero transitions between each), in practice there is a small set of sparse bosonic and spin-$s$ operators that are used most often. Here we summarize the set of $d$-level operators used in this study, where it is conceptually useful to explicitly write down some $d$-by-$d$ matrix representations.



Bosonic operators can be constructed from the well-known ladder operators $\hat a$ and $\hat a^\dag$, where (importantly for encoding considerations) all non-zero terms $|l \ra\la l'|$ obey $|l-l'|=1$. 
The position operator $\hat q=\frac{1}{\sqrt{2}} (\hat a_j^\dag + \hat a_j)$ is tridiagonal with zeros on the diagonal:

\begin{equation}
\begin{split}
\hat q = \frac{1}{\sqrt{2}}\begin{bmatrix}
0 & 1 & 0 & 0 & \dots \\
1 & 0 & \sqrt{2} & 0 & \dots \\
0 & \sqrt{2} & 0 & \sqrt{3} & \dots \\
0 & 0 & \sqrt{3} & 0 & \dots \\
\vdots & \vdots & \vdots & \vdots & \ddots
\end{bmatrix}
\end{split}
\end{equation}


This means that the square of $\hat q$, often used in vibrational and bosonic Hamiltonians, is pentadiagonal but with zeros for terms where $|l-l'|=1$:

\begin{equation}\label{eq:qSq}
\hat q^2 = \half
\begin{bmatrix}
 1 & 0 & \sqrt{1\cdot2} & 0 & 0 & \dots \\
 0 & 3 & 0 & \sqrt{2\cdot3} & 0 & \dots \\
 \sqrt{1\cdot2} & 0 & 5 & 0 & \sqrt{3\cdot4} & \dots \\
 0 & \sqrt{2\cdot3} & 0 & 7 & 0 & \dots \\
 0 & 0 & \sqrt{3\cdot4} & 0 & 9 & \dots \\
\vdots & \vdots & \vdots & \vdots & \vdots & \ddots
\end{bmatrix}.  
\end{equation}


Notably, $|l-l'|$ 
for non-zero entries is either 0 or 2, making the Gray code less useful for this operator. The momentum operator $\hat p=\frac{i}{\sqrt{2}} (\hat a_j^\dag - \hat a_j)$ and its square $\hat p^2$ have the same sparsity patterns as $\hat q$ and $\hat q^2$ respectively. The number operator $\hat n=\hat a^\dag \hat a$ of equation \eqref{eq:n_explicit} is diagonal and DBD, which leads to efficient SB mappings as discussed in Section~\ref{sec:dbd}. Finally, we study the two-site bosonic interaction operator 

\begin{equation}
\hat{a}_i ^\dag \hat{a}_{j} + \hat{a}_i \hat{a}_{j} ^\dag.
\end{equation}








In order to consider spin Hamiltonians such as 
Heisenberg models\cite{Levitt2008,lora_serrano_16}, we encode spin-$s$ operators of arbitrary $s$, where the number of levels is $d=2s+1$. Matrix elements for transitions $| l \ra\la l' |$ are defined as follows \cite{merzbacher04}
\begin{equation}\label{eq:sz}
 \la l | \Sz | l' \ra  = \hbar(s+1-l)\delta_{l,l'}
\end{equation}
\begin{equation}\label{eq:sx}
\la l | \Sx | l' \ra  = \frac{\hbar}{2}(\delta_{l,l'+1} + \delta_{l+1,l'}) \sqrt{(s+1)(l+l'-1)-ll'}
\end{equation}
\begin{equation}\label{eq:sy}
\la l | \Sy | l' \ra  = \frac{i\hbar}{2}(\delta_{l,l'+1} - \delta_{l+1,l'}) \sqrt{(s+1)(l+l'-1)-ll'}
\end{equation}
where $\delta_{\alpha,\beta}$ is the Kronecker delta. The $\Sz$ are DBD operators, while $\Sx$ and $\Sy$ are tridiagonal with zeros on the diagonal, the same sparsity pattern as bosonic $\hat p$ and $\hat q$ operators.


The local operators considered thus far are effectively second quantization operators---each ket tends to correspond to an eigenstate in an isolated $d$-level system. Also of note is a recently proposed approach \cite{macridin18a, macridin18b} which maps bosons to qubits using the first quantized representation of the quantum harmonic oscillator.
The original proposal maps Hermite-Gauss functions, the eigenfunctions of the quantum harmonic oscillator, into a discretized position space. The approximate position operator is defined as
\begin{equation}\label{eq:x_firstquant}
\tilde X_{\textrm{FQ}} = \sum_{i=0}^{N_x-1} x_i | x_i \ra\la x_i |
\end{equation}
and $N_x$ is the number of discrete position points such that

\begin{equation}
x_i = (i-N_x/2)\Delta, i \in [0,N_x-1]
\end{equation}
where $\Delta$ is chosen such that the desired highest-order Hermite-Gauss function is contained within $(x_0,x_{N_x-1})$. Advantages and disadvanteges are discussed in the Supplementary Section V. We raise the possibility of using this approach partly to point out that a $N_x$-by-$N_x$ matrix operator may be mapped to qubits using the exact same procedure as the other operators, with $N_x$ replacing $d$. Note that $\tilde X_{\textrm{FQ}}$ is a DBD operator.


Quantum circuits for approximating the exponential of each operator and for each $d$ were compiled and then optimized using the procedure given in the Supplementary Section II. The optimization consists of searching for and performing gate cancellations where possible. For instance, two adjacent CNOT gates or two adjacent Hadamard gates will cancel.
Entangling gate counts for the optimized circuits of bosonic operators $\hat q$, $\hat q^2$, and $\inttwo$ are plotted in Figure \ref{fig:loc_d8}. We place significant focus on smaller $d$ values because they tend to be more common in physics simulation, but we note that applications requiring larger $d$ values do exist, for example in vibronic simulations where occupation numbers can approach $d=70$ \cite{sawaya19}. 


Comparing the tridiagonal operator $\hat q$ with the upper bounds given in Figure \ref{fig:count_qub_op_ub} demonstrates that the circuit optimization greatly reduces gate count for the compact codes and block unary, often by a factor of 2 to 3. On the other hand, the unary encoding effectively sees no improvement from optimization, though it remains the code with fewest entangling gates for a large subset of $d$ values.

As was the case in the upper bound calculations, operators built from tridiagonal matrices ($\hat q$ and $\inttwo$) show the Gray encoding outperforming SB, though after optimization the advantage is less pronounced. In contrast, for the pentadiagonal $\hat q^2$, the Gray code outperforms SB asymptotically, while SB is better for lower $d$ values (and lower $d$ values are likely to be more common in relevant bosonic Hamiltonian simulations). The changed trend can be explained by noting that the unity Hamming distance of the Gray code is not as advantageous for the sparsity structure of $\hat q^2$, given in equation \eqref{eq:qSq}. Also notable is the apparent dip in operation count at $d$=8, due to the fact that the diagonal of $\hat q^2$ is DBD.

Importantly, when mapping bosonic problems using compact encodings, it is sometimes the case that increasing the truncation value $d$ is beneficial. For instance, suppose one knows one can safely truncate at $d=5$ for a bosonic problem. When implementing $\hat q^2$, one would instead simply implement the operator for $d=8$, as the number of gates decreases while the number of qubits remains the same. Note that this is not possible in spin-$s$ particles, as it would cause leakage to unphysical states.

One of the more intriguing results is that the unary code is often inferior to the Gray or SB encodings. Pronounced examples of this inversion include $d$=4,7,8 for $\hat q$ and $d$=4 and 8 for $\hat q^2$, among others. This is notable because, for these values of $d$, Trotterizing the operator requires \textit{both} fewer qubit and fewer operations if Gray or SB is used. These results are in contrast to the naively expected trend that there would be a more consistent trade-off between qubit count and operation count.

Results for single-particle spin-$s$ operators $\hat S_x$ and $\hat S_z$, as well as interaction operator $\Sint$ are plotted in Figure \ref{fig:locspin}. Unlike bosonic Hamiltonians, the $d$ values are not simulation parameters but are determined by $s$ in the system we wish to simulate. The trends in spin operators tend to be more unruly than those in the bosonic operators.

Analogous to the bosonic case, $\hat S_z$ is DBD and therefore SB requires only single-qubit gates when $d$=4,8 ($s$=$\frac{3}{2}$,$\frac{7}{2}$) and no entangling gates. For these two values, SB uses both the fewest operations and the fewest qubits (fewer than unary). However, the Gray code is superior to SB for other values of $s$, the same behavior seen in the general DES matrix of Figure \ref{fig:dbd}. Because $\Sz$ is diagonal, the unary always requires just $d$ single-qubit rotations and no entangling gates. The two-particle operator $\Sint$ displays similar trends to $\Sz$.

As expected, the Gray code is usually superior to SB for the tri-diagonal $\Sx$, because of the unity Hamming distance between nearest levels. Unary is inferior in both gate count and qubit count for most values, a result highlighted earlier in low-$d$ bosonic operators.

For both bosonic and spin-$s$ operators, we have until now omitted discussion of the Gray-based block unary encoding with parameter $g=3$. There is never a case where this BU mapping is the sole encoding with the lowest entangling gate count. However, at least in principle, there may be limited cases where a particular hardware budget (Figure \ref{fig:hardware_budget}) dictates the need for a block unary encoding. A necessary condition for even considering the use of BU$^{Gray}_{g=3}$ is that its operation count is less than both compact codes, but more than unary. 
In such cases ($\hat q$ and $\inttwo$ for $d$=9; $\hat n^2$ $\sim \hat O^2$ in Figure \ref{fig:dbd} for several values; $\Sint$ and $\Sz$ for $s$=2), there may be a particular hardware budget would require this encoding for its particular memory/operation trade-off. Such highly specific hardware budgets seem unlikely to often appear. 




Note that the results herein should generally be considered constant-factor savings, because in most relevant systems $d$ does not increase with system size, \textit{i.e.} with the number of particles. For the simulation of scientifically relevant quantum systems, Hamiltonians are composed of more than one simple operator. For such situations, one may calculate the overall cost within a given encoding, as we do in Section \ref{sec:physical-systems}. As will be discussed in Section \ref{sec:conversion}, it is often beneficial to Trotterize different parts of the Hamiltonian in different encodings, if the cost difference outweighs the overhead of conversion.







\subsection{Conversions between encodings}
\label{sec:conversion}
\begin{figure}[htb]
    \centering
    \includegraphics[width=1.0\textwidth]{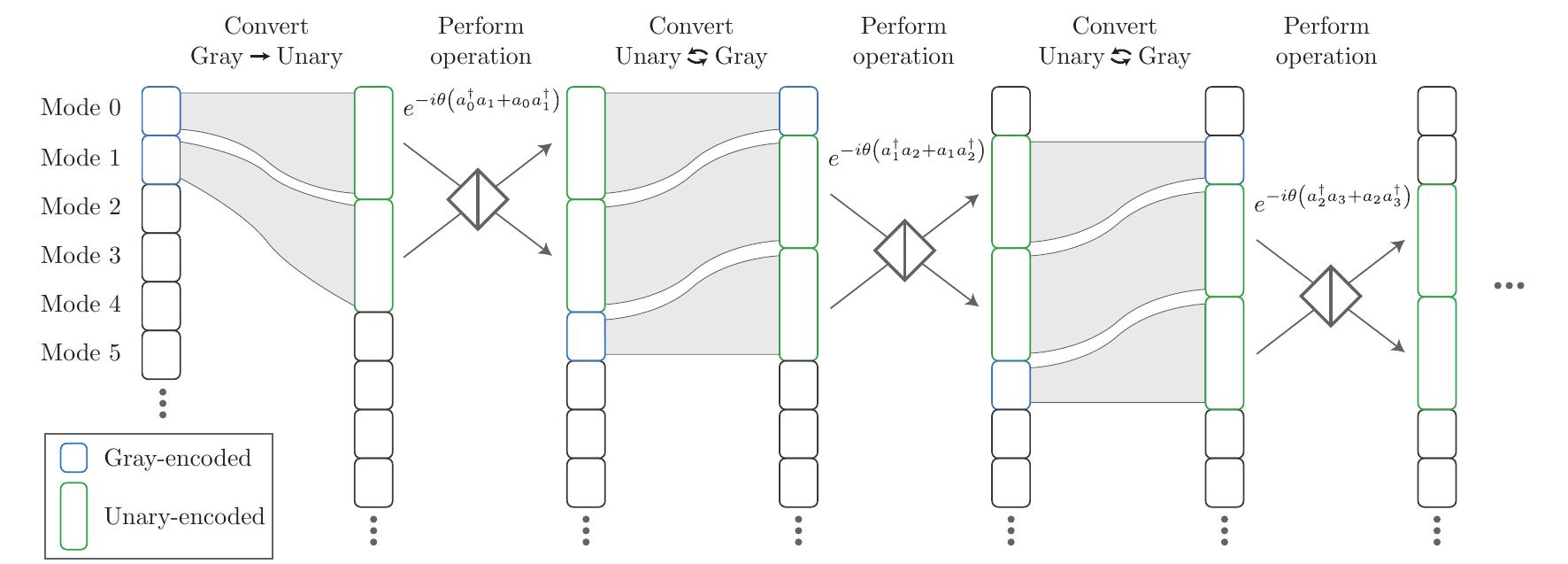}
    \caption{Schematic showing the utility of swapping between encodings. When extra memory resources are available and the unary code is the most efficient for implementing an operator, one may expand into the unary representation, perform the operation, and then compact the data back to SB or Gray. The example shown here is the bosonic interaction operator $\inttwo$. This operator is present in bosonic Hamiltonians and for digital simulations of beamsplitters. For many values of $d$, implementing this operator in unary is much cheaper than implementing it in Gray or SB.  When this strategy is worth the cost of conversion, the Hamiltonian simulation is in \textit{Scenario C} discussed in Section \ref{sec:physical-systems}. Hence each particle starts with $\ceil{\log_2 d}$ qubits, expands out to $d$ qubits, and then compacts back. Whether this procedure leads to cost savings is heavily dependent on the problem and the parameters.}
    \label{fig:schem_conv_enc}
\end{figure}


It is often the case that different terms in a Hamiltonian are more efficiently simulated in different encodings. For example, in the Bose-Hubbard model, the number operator $\hat n$ is usually more efficient in SB, while the hopping term $\hat{b}_i ^\dag \hat{b}_{i+1} + \textrm{h.c.}$ is usually more efficient in the Gray encoding (see Figure \ref{fig:loc_d8}). Here we show that the cost of converting from one encoding to another is often substantially less than the difference in resource efficiency between two encodings, which means that it can be advantageous to continually be compacting and uncompacting the data. For example, if unary is the most efficient for implementing an operator, one may wish to compact the data between operations to save memory resources, as shown in Figure \ref{fig:schem_conv_enc}. In this section we give general quantum circuits and resource counts for converting between all encodings considered in this work.



\begin{figure}[htb]
\centering
\includegraphics[width=0.7\textwidth]{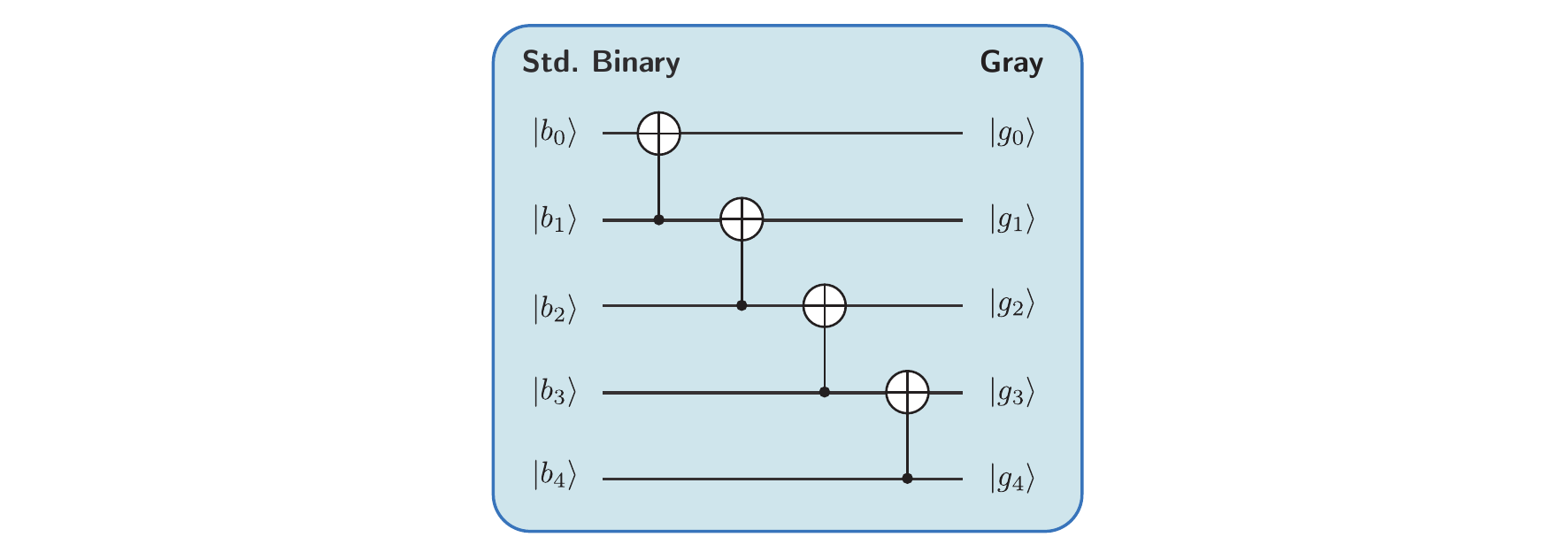}
\caption{Quantum circuit for SB to Gray conversion.}
\label{fig:binary-to-gray-circuit}
\end{figure}

One can convert between the Gray and SB encodings by applying $(\ceil{\log_2 d}-1)$ CNOTs in sequential order \cite{shukla15} as shown in Figure~\ref{fig:binary-to-gray-circuit}.


\begin{figure}[b]
\centering
\includegraphics[width=1.0\textwidth]{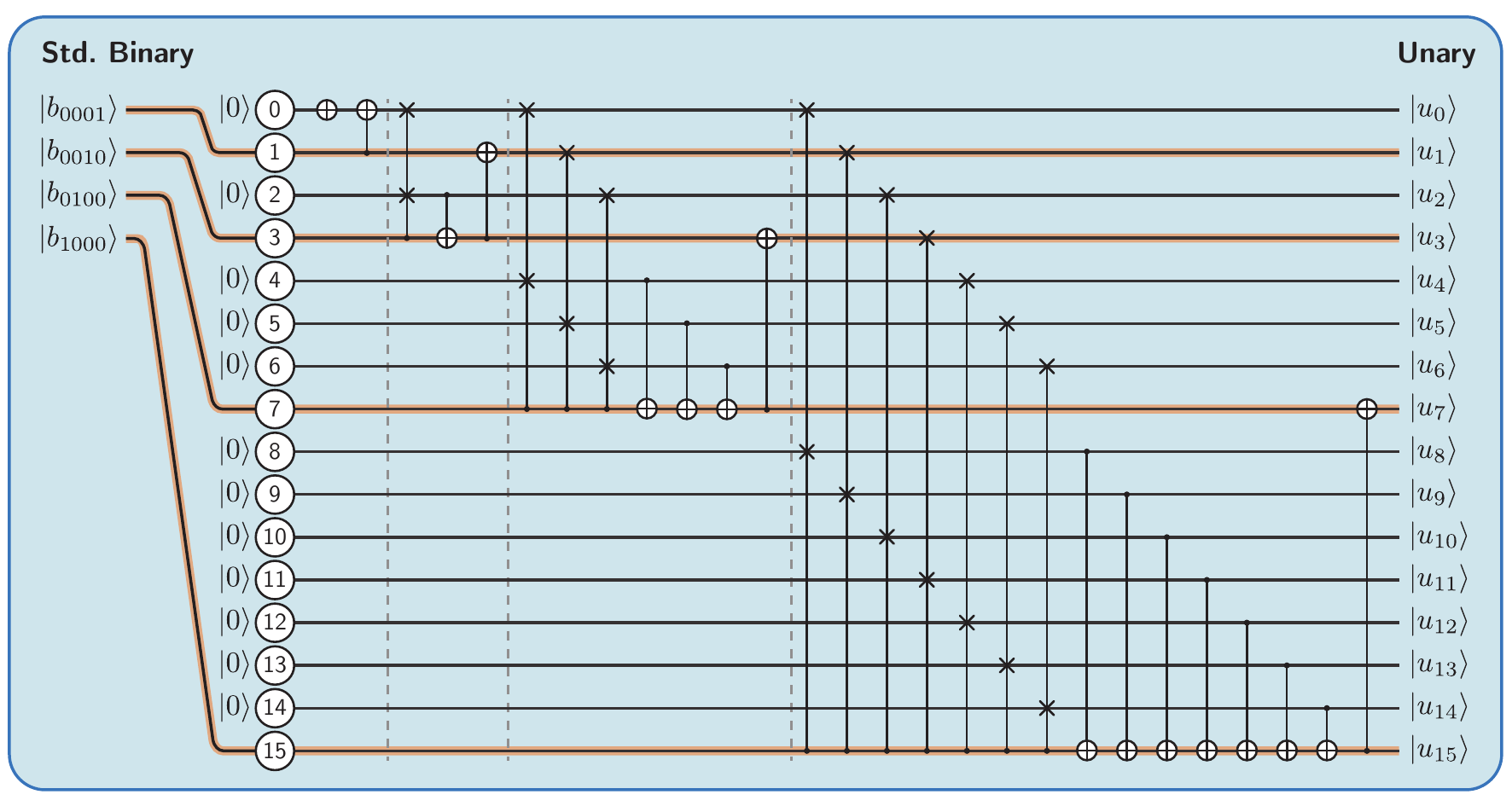}
\caption{Circuit to convert the SB representation into the unary representation.
Every CSWAP is accompanied by a CNOT. 
Modifications are required when $d$ is not a power of two, as discussed in the main text and Supplementary Section III.}
\label{fig:binary-to-unary-circuit}
\end{figure}

\begin{algorithm}[H]
\caption{An algorithm to build a quantum circuit for SB--unary conversion given an arbitrary truncation $d$. For multi-qubit gates, the first argument specifies the control qubit and the successive ones are targets.}
\label{alg:stdbin-unary-conversion}

\begin{algorithmic}
\Function{stdbin-unary-converter}{d}

\State $K \gets \ceil{\log_2 d}$ \Comment{Qubit count in SB}

\State SWAP($K-1$,$d-1$)  \Comment{Move positions of SB bits}
\For{$b \gets K-2$ to 0}
\State SWAP($b$,$2^{b+1}-1$)
\EndFor

\State X(0)  \Comment{First two gates}

\State CNOT(1,0)

\For{$b \gets$ 1 to $K-1$} \Comment{SWAP and CNOT gates}

\State $i_{ctr}$ = MININUM($2^{b+1} - 1$,$d-1$) \Comment{Define `control' bit}

\For{$L \gets$ $2^{b}$ to $(i_{ctr}-1)$}  
\State CSWAP($i_{ctr}$,$L$,$L-2^{b}$)
\EndFor

\For{$L \gets$ $2^{b}$ to $(i_{ctr}-1)$}   
\State CNOT($L$,$i_{ctr}$)
\EndFor

\State CNOT( $i_{ctr}$, $i_{ctr}-2^{b}$ ) \Comment{Last CNOT gate}

\EndFor

\EndFunction
\end{algorithmic}
\end{algorithm}

The conversion between unary and SB is more complex. The conversion may be especially relevant in a future fault-tolerant quantum computing era, when extra quantum resources are available, because the unary encoding becomes more beneficial as $d$ increases and because the conversion cost is significant. 
Inspired by previous work \cite{unary_online}, in Figure~\ref{fig:binary-to-unary-circuit} we show an example case for converting from SB to unary when $d=16$. 
A state is initially encoded in SB using qubits on the left, and the memory space is enlarged to include the number of qubits needed for unary. No ancilla qubits are required.
As quantum circuits are reversible, unary-to-SB conversion follows by inversion of the circuit. In Table \ref{tbl:stdbin-unary}, we provide the converter circuit resource count for a general $d$-level truncated quantum system, following Figure \ref{fig:binary-to-unary-circuit}. Resource counts assume a decomposition of CSWAP into Clifford+T gates \cite{kim2018efficient}. A general algorithm to build the converter circuit can be seen in the Algorithm \ref{alg:stdbin-unary-conversion}.
$\ceil{\cdot}$ and $\floor{\cdot}$ are respectively the ceiling and floor functions. The validity of the conversion procedure is most easily shown by tracing a single unary state through the reverse algorithm.
When $d$ is not a power of 2, modifications are needed. These modifications are already accounted for in Algorithm \ref{alg:stdbin-unary-conversion} and example circuits for $d=$5 and 7 are given in the Supplementary Section III.


\begin{table}[t]
\centering
\begin{tabular}{ p{4.1cm} p{5.3cm} }
 \hline \hline
 \multicolumn{1}{c}{Before Clifford+T decomposition} &  \multicolumn{1}{c}{After Clifford+T decomposition}  \\
 \hline 
$n(\textrm{CNOT}) = d - 1$ &  $n(\textrm{CNOT}) =  9d - 8\lceil \log_2 d \rceil - 9$\\
$n(\textrm{CSWAP}) = d - \lceil \log_2 d \rceil -1$ & $n(\textrm{H})      =  2d - 2\lceil \log_2 d \rceil - 2$\\
$n(\sx) = 1$ & $n(\sx) = 1$\\
& $n(\textrm{T})      =  4d - 4\lceil \log_2 d \rceil - 4$\\
& $n(\textrm{T}^\dag) =  3d - 3\lceil \log_2 d \rceil - 3$\\
 \hline
\end{tabular}
\caption{Resource counts for the SB to unary conversion, for arbitrary $d$.}
\label{tbl:stdbin-unary}
\end{table}

For completeness, we constructed a circuit for converting between SB and block unary, for BU$^{\textrm{SB}}_{g=3}$, shown in the Supplementary Section IV. We do not further analyze BU conversions, as block unary is expected to have limited utility, and even then it will usually be the case that decoherence times are too low to allow for conversions (see Section \ref{sec:operators}).



\subsection{Composite Systems}
\label{sec:physical-systems}



Here we consider resource counts for simulating five physically and chemically relevant Hamiltonian systems. The Hamiltonians correspond to the shifted one-dimensional QHO, the Bose-Hubbard model \cite{fisher89,bh_phases_1,bloch_review}, multidimensional molecular Franck-Condon factors \cite{huh15,mcardle19,sawaya19}, a spin-$s$ transverse-field Heisenberg model \cite{Levitt2008,lora_serrano_16}, and simulating Boson sampling \cite{aaronson14} on a digital quantum computer. The former four systems consist of an arbitrary number of $d$-level particles. For the Franck-Condon factors, the Duschinsky matrix is assumed to have a constant $k=4$ nonzero entries per row. With the exception of the simple QHO, all of these problem classes would benefit from digital quantum simulation, because there are limits to the theoretical and practical questions that can be answered by classical computers. Supplementary Section VI gives a more thorough overview of these problems.
Assuming that $d$ remains constant as the particle number increases, differences in resource counts between mappings are constant-factor savings that are independent of system size.

Using resource counts from the optimized circuits for Trotterizing individual operators and from the circuits for inter-encoding conversion, we calculated and compared the required two-qubit entangling gate counts for the selected composite Hamiltonians. We considered five encoding schemes: (i) SB only, (ii) Gray code only, (iii) unary only, (iv) allowing for conversion between SB and Gray, and (v) using all three while compacting to save memory. For (iv) and (v), the reported results include the cost of conversion. To the best of our knowledge, schemes (ii), (iv), and (v) are novel to digital quantum simulation. In Figure \ref{fig:schem_conv_enc}, an example of encoding scheme (v) is shown. These encoding schemes do \textit{not} directly correspond to the `scenarios' discussed below; the scenarios denote the optimal encoding scheme under different hardware budgets.

The result for (iv), the encoding scheme that combines both the SB and Gray codes, is reported only when it represents an improvement over both SB-only and Gray-only. We give results for (v), which compacts and uncompacts the qubits for unary computations, only when the unary code was the most efficient of the first four encoding schemes. We give all results in terms of resource counts relative to the SB mapping, noting again that the relative resource requirements between encoding schemes are independent of system size (\textit{i.e.} number of particles or modes).

For some local bosonic operators, the number of entangling gates is not a monotonically increasing function of $d$. Such operators include $\hat n$, $\hat n^2$, and $\hat q^2$ (Figures \ref{fig:dbd} and \ref{fig:loc_d8}). Our numerics for the composite systems take this into account, increasing the cutoff $d$ if it is beneficial. For instance, if $d=5$ is a sufficient truncation and we implement $\hat q^2$, we use resource counts for $d=8$, because this uses the same number of qubits but fewer operations (Figure \ref{fig:loc_d8}). This trick is not possible for the spin-$s$ systems, where $d$ is determined not by a sufficient truncation value but by the nature of the particle itself (its spin $s$).

A selection of resource comparisons is shown in Figure \ref{fig:composites}. We show results from $d=4$ and 10 because they highlight the variety of rankings that occur, and demonstrate that the best encoding scheme can be highly sensitive to $d$ even within the same Hamiltonian class. Numerical results up to $d=16$ are given in the Supplementary Section VII. 

In terms of which encoding class should be used, the results can be categorized into four scenarios. \textit{Scenario A} applies when using just one of the compact encodings (SB or Gray) is the best choice. Of the results shown in Figure \ref{fig:composites}, the Bose-Hubbard and 1D QHO models for $d=4$ fit this description. The optimal choice is to stay in one of the compact encodings for the entire calculation, while never using more than $\log_2 d$ qubits per particle for the calculation.

\textit{Scenario B} refers to Hamiltonians for which the optimal strategy is to use a compact amount of memory but to allow for conversion between Gray and SB. This includes the Heisenberg model for $s=\frac{7}{2}$ and the Franck-Condon Hamiltonian for $d=4$. Because the cost of conversion is very small, this scenario usually implies that at least one of the local operators in the Hamiltonian are optimal in Gray, at least one is optimal in SB, and none are optimal in unary. To take the Heisenberg model with $s=\frac{7}{2}$ as an example, one can see this is the case by comparing $\Sx$ and $\Sint$ in Figure \ref{fig:locspin}.

\textit{Scenario C} applies when unary is the superior encoding \textit{and} it is still considered the best encoding even if one repeatedly unravels and compacts to preserve memory (as in Figure \ref{fig:schem_conv_enc}). This latter trait is important because it means that, even including the substantial cost of SB-to-unary conversion, with a cost of approximately $9d$ entangling gates (Table \ref{tbl:stdbin-unary}), it is still better to convert back and forth between unary and SB/Gray. This is true even when memory constraints require that one stores the information compactly for most of the time. This occurs with $d=10$ for the Bose-Hubbard, boson sampling, and Franck-Condon Hamiltonians, all of which are bosonic problems. 

\textit{Scenario D} refers to cases where unary is the superior encoding, assuming that the information is \textit{not} compacted back to SB in order to save memory. This scenario implies that, if one has the qubit space to stay in unary form for the entire calculation, unary is optimal. If one does not have the memory resources for this, it is best to simply perform all operations in Gray and/or SB. The reason for this discrepancy is that the cost of converting binary to unary is substantial, as mentioned above. This scenario applies to the 1D QHO for $d=10$ as well as the Heisenberg model for $s=2$.

Results for $d$ values up to 16 are given in the Supplementary Section VII. Note that the novel memory-efficient schemes (ii), (iv), and (v) did not lead to improvements in every case; in a minority of the problem instances we considered, the optimal encoding scheme was to use only SB or only unary. By memory-efficient we mean that the scheme stores the encoded subsystems in $\ceil{\log_2 d}$ qubits, with the possible exception of when they are being operated on. Comparing to the memory-\textit{in}efficient unary-only scheme, our novel approaches reduced two-qubit entangling gate counts by up to 33\%. Compared to the memory-efficient SB-only scheme, we observed gate count reductions of up to 49\%. The latter case is more relevant when qubit count is a substantial constraint. These savings are especially important for running algorithms on near-term hardware, since by simply modifying the encoding procedure one can substantially decrease the effective circuit depth.

\section{Discussion}
\label{sec:conclusions}

After introducing a general framework for encoding $d$-level systems to multi-qubit operators, we have analyzed the utility and trade-offs of several integer-to-bit encodings for qubit-based Hamiltonian simulation. The mappings may be used for Hamiltonians built from subsystems of bosons, spin-$s$ particles, molecular electronic energy levels, molecular vibrational modes, or other $d$-level subsystems. 
We analyzed the mappings primarily in terms of qubit counts and the number of entangling operations required to estimate the exponential of an operator.



Of the Gray and SB codes, we demonstrated that the Gray code tends to be more efficient for tridiagonal matrix operators, while SB tends to be superior for a common class of diagonal matrix operator. Importantly, we show that converting between encodings within a Suzuki-Trotter step often leads to savings. Notably, though the unary code tends to require more qubits but fewer operations, it is often the case that the SB or Gray code is more efficient \textit{both} in terms of qubit counts and operation counts. To the best of our knowledge, the Gray code had not been previously used in Hamiltonian simulation.

We compared resource requirements between encodings for the following composite Hamiltonians: the Bose-Hubbard model, one-dimensional quantum harmonic oscillator, vibronic molecular Hamiltonian (\textit{i.e.} Franck-Condon factors), spin-$s$ Heisenberg model, and boson sampling. The optimal encoding, and whether it was beneficial to interconvert between encodings, was heavily dependent both on the Hamiltonian class and on the truncation level $d$ for the particle.
We placed optimal encoding strategies into four different ``scenarios,'' each of which points to a different optimal encoding and simulation strategy. The simulation scenario depends on which encodings require the fewest operations, on whether interconverting between mappings is worth the additional cost, and on qubit memory constraints. The many anomalies in our results highlight the need to perform an analysis of each new class of Hamiltonian simulation problem, determining numerically which simulation strategy is optimal before performing a simulation on real hardware.

There are several directions open for future research. First, there are ways to analyze resource requirements other than enumerating the entangling operations. For long-term error-corrected hardware, estimating T gate count may be most relevant \cite{gosset13}. Additionally, we assumed all-to-all connectivity in this work, which tends to be a feature of ion trap quantum computers \cite{figgatt19}.
But other quantum hardware types require one to consider the topology of the qubit connections and implementation of SWAP gates \cite{holmes18}, a consideration that would modify the resource counts and may modify some trends observed here.

We envision that the methodology and results of this work will be helpful for both theorists and experimentalists in designing resource efficient approaches to quantum simulation of a broader set of physically and chemically relevant Hamiltonians.

%
%

\section*{Data availability:} The data that support the findings of this study are available upon reasonable request.

\section*{Acknowledgements}

A.A.-G., T. H. K. and  T. M. acknowledge funding from Intel  Research and Dr. Anders G. Frøseth. A.A.-G. also acknowledges support from the Vannevar Bush Faculty Fellowship and the Canada 150  Research Chairs Program. The authors declare that there are no competing interests.

\section*{Author Contributions}
N.P.D.S. conceived of and designed the study, developed the theory, and generated and analyzed the data. 
T.M. developed encoding conversion schemes, analyzed data, and created conceptual figures. 
T.H.K. developed encoding conversion schemes. 
S.J. optimized all quantum circuits. 
A.A.-G. supervised work on encoding conversions. 
G.G.G. conceived of the encoding conversions section and developed theory. 
N.P.D.S., G.G.G., T.M., and T.H.K. wrote the manuscript. 
All authors edited the manuscript.

\section*{Competing Interests}
The authors declare no competing interests.


\bibliography{refs}


\appendix

\section*{Figure Legends}

\textbf{FIG. 1.} Especially for near-term noisy quantum hardware, gate counts and qubit counts will be limited. In principle, these constraints can be used to approximate a \textit{hardware budget} for a set of hardware and a particular Hamiltonian simulation problem. For example, if one wants to simulate a collection of $N$ bosons on a small quantum computer, the decoherence time and gate errors will constrain the allowed number of gates, while the total number of qubits will constrain the qubit count per boson. In this schematic, we show two arbitrary hardware budgets for Trotterizing the exponential of $\hat q^2$ for one boson with truncation $d=5$. In device A, both the Gray and standard binary encodings are satisfactory, but the unary code requires too many qubits. However, because device B allows for more qubits but fewer operations, the unary code is sufficient while the former two encodings require too many operations. This highlights the need for considering multiple encodings, as an encoding that is best for one type of hardware is not necessarily universally superior.

\textbf{FIG. 2.} Using an arbitrary selection of parameters for common physics and chemistry Hamiltonians, we have plotted the comparative computational costs required for first-order Trotterization. Costs are reported in terms of number of two-qubit entangling gates, \textit{relative to} the cost of standard binary (SB). The three encodings shown here---standard binary, Gray code, and unary---are defined in the text. The five Hamiltonians are the Bose-Hubbard model, one-dimensional quantum harmonic oscillator (QHO), Franck-Condon calculation, boson sampling, and spin-$s$ Heisenberg model. The optimal encodings are sensitive both to the Hamiltonian class and the number of levels $d$ (determined by bosonic truncation or by the spin value $s$).
In some cases, it is best to stay in a particular encoding for the duration of the simulation.
Other times, it is worth bearing the resource cost of converting between encodings, because it saves on total operations. Still other times, the decision to save operations by converting between encodings will depend on whether available hardware is gate count limited or qubit count limited. Four Scenarios, A through D, are discussed in Section \ref{sec:physical-systems}.

\textbf{FIG. 3.} Canonical quantum circuits used to exponentiate Pauli strings on a universal quantum computer. One needs $2(p-1)$ two-qubit gates for such an operation, where $p$ is the number of Pauli operators in the term.  
When a product of many exponentials is used, as in the Suzuki-Trotter procedure, there tends to be significant gate cancellation.

\textbf{FIG. 4.} Numerical upper bounds for resource counts of implementing one Suzuki-Trotter step of a $d$-by-$d$ real Hermitian matrix operator $\hat B$, where $\hat B$ is tridiagonal with zeros on the diagonal. Top: Qubit counts for mappings considered in this work. BU$^{Gray}_{g}$ stands for block unary where $g$ is the size of the block. Asymptotically, the number of qubits scales logarithmically for the SB and Gray encodings, and linearly for the unary and block unary encodings. Bottom: Upper bounds of CNOT operation counts for implementing one Suzuki-Trotter step of $\hat B$. This is the sparsity pattern of canonical bosonic position and momentum operators as well as the $S_x$ spin operators in spin-$s$ systems. Upper bounds were calculated by mapping the full operator to a sum of weighted Pauli strings, combining terms, and then using equation \eqref{eq:ncxub_sgl_pauli}. Notably,  encodings with higher qubit counts tend to have lower upper bounds for gate counts, and vice-versa.

\textbf{FIG. 5.} Entangling gates for a single Suzuki-Trotter step of an arbitrary diagonal evenly-spaces (DES) operator $\hat O$ (left), and its square (right). Gate counts are for optimized quantum circuits. DES operators are a subset of the diagonal binary-decomposable (DBD) operator class. Because it is diagonal, the unary code always requires only single-qubit operations. When $\log_2 d$ is an integer, the SB code requires no entangling gates and just $\log_2 d$ single-qubit operations, making it the most efficient encoding (in terms of both qubits and operations). DBD operators are a common operator class, encompassing \textit{e.g.} the bosonic number operator $\hat n$ and the spin-$s$ operator $\hat S_z$.

\textbf{FIG. 6.} CNOT gate counts for optimized circuits of the bosonic position operator (first row), position operator squared (second row), and two-site bosonic interaction operator (third row). Plots on the left correspond to enlargements of the dotted boxes in the plots on the right. There are several notable trends and anomalies: (a) Though unary usually requires the fewest operations as $d$ increases, there are several operators for which the Gray or SB code is more efficient than the unary in \textit{both} qubit and operation count. This occurs most pronouncedly at values such as $d=4$, 7, and 8. (b) The Gray code is usually more efficient than SB even after circuit optimization, especially for operators composed of tridiagonal operators (top and bottom rows). (c) The Gray code's advantage is either less pronounced or disappears for $\hat q^2$, because $\hat q^2$ is a pentadiagonal operator, for which the unity Hamming distance of the Gray code is less useful. (d) The reduction in operation count for $\hat q^2$ at values such as $d=8$, 24, or 32 occur because the diagonal of $\hat q^2$ is DBD.

\textbf{FIG. 7.} CNOT gate counts of optimized Suzuki-Trotter circuits, for approximating the exponentials of the spin-$s$ operators shown. There is no clean overall trend for $\Sx$ and $\Sint$ (except that Gray tends to out-perform SB), highlighting the need to study encodings thoroughly for each new use case. Notably, because $\hat S_z$ is diagonal binary-decomposable, for values of $s= \frac{3}{2} , \frac{7}{2} $ (which are 4- and 8-level system respectively) the SB code requires both fewest operations and fewest qubits. 

\textbf{FIG. 8.} Schematic showing the utility of swapping between encodings. When extra memory resources are available and the unary code is the most efficient for implementing an operator, one may expand into the unary representation, perform the operation, and then compact the data back to SB or Gray. The example shown here is the bosonic interaction operator $\inttwo$. This operator is present in bosonic Hamiltonians and for digital simulations of beamsplitters. For many values of $d$, implementing this operator in unary is much cheaper than implementing it in Gray or SB.  When this strategy is worth the cost of conversion, the Hamiltonian simulation is in \textit{Scenario C} discussed in Section \ref{sec:physical-systems}. Hence each particle starts with $\ceil{\log_2 d}$ qubits, expands out to $d$ qubits, and then compacts back. Whether this procedure leads to cost savings is heavily dependent on the problem and the parameters.

\textbf{FIG. 9.} Quantum circuit for SB to Gray conversion.

\textbf{FIG. 10.} Circuit to convert the SB representation into the unary representation.
Every CSWAP is accompanied by a CNOT.
Modifications are required when $d$ is not a power of two, as discussed in the main text and Supplementary Section III.


\end{document}


\title{SUPPORTING INFORMATION: Resource-efficient digital quantum simulation of $d$-level systems for photonic, vibrational, and spin-$s$ Hamiltonians.}

\author{Nicolas P. D. Sawaya}
\email{nicolas.sawaya@intel.com}
\affiliation{\intelcali}
\author{Tim Menke}
\affiliation{Department of Physics, Harvard University, Cambridge, MA 02138, USA}
\affiliation{Research Laboratory of Electronics, Massachusetts Institute of Technology, Cambridge, MA 02139, USA}
\affiliation{Department of Physics, Massachusetts Institute of Technology, Cambridge, MA 02139, USA}
\author{Thi Ha Kyaw}
\affiliation{\utcomp}
\affiliation{\utchem}\
\author{Sonika Johri}
\affiliation{\inteloregon}
\author{Al\'an Aspuru-Guzik}
\affiliation{\utcomp}
\affiliation{\utchem}
\affiliation{\vectorinst}
\affiliation{\cifar}
\author{Gian Giacomo Guerreschi}
\email{gian.giacomo.guerreschi@intel.com}
\affiliation{\intelcali}





\maketitle

\tableofcontents



\section{Upper bounds for entangling gates}

The number of required quantum operations is closely related to the lengths of the Pauli strings in the encoded Hamiltonian. The distribution of lengths of Pauli strings may be analyzed in terms of binomial distributions, because the quantity to be analyzed is a product of two-term sums, \textit{i.e.} each $| x_i \ra\la x'_i |_i$ in the expression
\begin{equation}\label{eq:ll_to_xx}
| l \ra\la l' | \mapsto \bigotimes_{i \in C(l) \cup C(l')} | x_i \ra\la x'_i |_i
\end{equation}
is mapped to a sum of exactly two one-qubit unitary operators, as outlined in the main text.
Consequently, for a diagonal term $| l \ra\la l |$ the number of occurrences of length-$p$ Pauli strings is
\begin{equation}\label{eq:f_binom_diagterm}
f_{|l \ra\la l|}(p; K) = {K \choose p},
\end{equation}
where $K$ is $|C^{\textrm{enc}}(l) \cup C^{\textrm{enc}}(l')|$, \textit{i.e.} the number of qubits in the relevant bitmask subsets. This is because every term in equation \eqref{eq:ll_to_xx} is $|0\ra\la0|$ or $|1\ra\la1|$, which respectively map to $\half (I + \sz)$ and $\half (I - \sz)$. Hence all terms in the product have a single identity---consider \textit{e.g.} $(I + \sz)(I - \sz)\cdots(I + \sz)$---which leads to the Pauli string lengths having a binomial distribution.

We consider a Hermitian term $\alpha\llp$ + $\alpha^*\lpl$ where $l \neq l'$. For such a term, the resulting Pauli operator for this simple term pair will also be related to the binomial distribution. The number $f$ of occurrences of length-$p$ Pauli strings is
\begin{equation}\label{eq:f_binom}
f(p; d_H,K) = \half 2^{d_H} {K-d_H \choose p-d_H},
\end{equation}
where $d_H=d_H(\mathcal{R}(l),\mathcal{R}(l'))$ and $\mathcal R$ depends on the chosen encoding.  The reason for the $\half$ factor is that non-Hermitian terms (\textit{i.e.} Pauli strings with an imaginary coefficient) cancel out when we consider the sum $\alpha\llp$ + $\alpha^*\lpl$. The non-Hermitian term $\llp$ would not have the $\half$ factor when considering the distribution of Pauli string lengths, but we do not analyze such terms because the goal is to exponentiate Hermitian operators using quantum circuits.

It is conceptually useful to consider the case of all bits being mismatched, \textit{i.e.} when $d_H=K$ for a term $\alpha\llp$ + $\alpha^*\lpl$. In this case, because there are no identity matrices present in the original product of equation \eqref{eq:ll_to_xx}, there are $\half 2^{d_H}{K \choose 0}=2^{d_H-1}$ Pauli strings, all with length $K$.

Combining equation \eqref{eq:f_binom} with $n_{\textrm{cx}}(p) = 2(p-1)$ from equation (19) in the main text, an upper bound (denoted \textit{UB}) for the number of CNOT gates is therefore

\begin{equation}\label{eq:2lev_cx_ub}
\begin{split}
n_{\textrm{cx,UB}}[d_H(l,l'),K] = \sum_{p=2}^{K}f(p;d_H,K)(2p-2) \\
= \half \sum_{p=2}^{K} 2^{d_H} {K-d_H \choose p-d_H} (2p-2),
\end{split}
\end{equation}

where the sum starts at $p=2$ because CNOT gates are required only for $p>1$. This is an upper bound because gate cancellations are possible once the circuit has been constructed.

Using known binomial identities given below, and again considering only the term $\alpha\llp$ + $\alpha^*\lpl$ or the diagonal $|l \ra\la l|$, this becomes (duplicated in the main text)

\begin{equation}\label{eq:ncxub_binom_gen}
\begin{split}
n_{\textrm{cx,UB}}[d_H=0,K] = (2^K K-2(2^K) - 2), \\
n_{\textrm{cx,UB}}[d_H=1,K] = \half (2^K K-2^K),  \\
n_{\textrm{cx,UB}}[d_H=2,K] = \half (2^K K). \\
\end{split}
\end{equation}

The special case of $d_H=0$ represents a diagonal term $|l \ra\la l|$. In the case of compact codes (standard binary and Gray) for which $\log_2 d$ is an integer, these formulas can be expressed in terms of the number of levels $d$ as
\begin{equation}\label{eq:ncxub_binom_dense}
\begin{split}
n_{\textrm{cx,UB}}[d_H=0,K] = d\log_2 d - 2d - 2,  \\
n_{\textrm{cx,UB}}[d_H=1,K] = \half (d\log_2 d - 2d ), \\
n_{\textrm{cx,UB}}[d_H=2,K] = \half d\log_2 d.  \\
\end{split}
\end{equation}
We derived formulas \eqref{eq:ncxub_binom_gen} from the following binomial identities

\begin{equation}
\sum_j^K {K \choose j} = 2^K
\end{equation}

\begin{equation}
{K \choose j} = \frac{K}{j} {K-1 \choose j-1}
\end{equation}

\begin{equation}
{K \choose j} = \frac{K(K-1)}{j(j-1)} {K-2 \choose j-2}.
\end{equation}
Again, thus far we have considered upper bounds only for one term $\alpha\llp$ + $\alpha^*\lpl$ or one element $\lkbl$.

We now consider upper bounds for typical full matrix operators using a compact code. Equation \eqref{eq:ncxub_binom_dense} implies that for each term $\alpha\llp$ + $\alpha^*\lpl$, $n_{\textrm{cx,UB}}$ is of order $O(d \log d)$.
Hence, for the sparsity pattern of $\hat B$ discussed in the main text---for which the number of non-zero entries in the operator is $O(d)$---the compact encodings (Gray and SB) yield an upper bound of $O(d)O(d\log d)$=$O(d^2\log d)$ entangling gates.

On the other hand, equation \eqref{eq:ncxub_binom_gen} shows that for the unary encoding, one term pair $\alpha\llp$ + $\alpha^*\lpl$ has $n_{\textrm{cx,UB}} \sim$ $O(1)$, since the number of CNOT gates is independent of $d$.
For sparsity pattern $\hat B$ this gives upper bounds of $n_{\textrm{cx,UB}} \sim $ $O(d)$, substantially less asymptotically than the compact encodings. This improved bound comes at the cost of the unary and BU encodings requiring more qubits. 

For an operator 
corresponding to a \textit{dense} matrix of $O(d^2)$ nonzero entries and represented with a compact encoding, the worst-case scaling would be $O(d^2)O(d\log d)$=$O(d^3 \log d)$, but this is clearly an overestimate.
In fact, there are only $O(4^{\log_2 d})$ = $O((2^{\log_2 d})^2)$ = $O(d^2)$ distinct Pauli strings for $\log_2 d$ qubits and each of them requires at most $O(\log d)$ CNOT gates, for a total of $O(d^2 \log d)$ entangling gates (this is the absolute limit based on counting and implementing all possible Pauli operators of length $\log_2 d$).
%
As noted in the main text, the result is that the unary code has a diminished advantage as the density of the matrix operator increases.

Finally, in the block unary encoding for one Hermitian pair, the bitmask subset union has $K \sim O(\log g)$ bits, resulting in $O(g \log g)$ operations, in analogy to the compact case. Hence, sparsity pattern $\hat B$ yields a bound of $O(d g \log g)$ and a dense matrix yields $O(d^2 \log g)$.

In many cases these analytical expressions will not be particularly useful in calculating which encoding is best for a given operator, except to explain general trends. 
When including all terms in the matrix operator, there will be many repeated Pauli matrices whose coefficients will be summed, often leading to a reduction in the number of terms.
Circuit optimizers, like the one used in this work, should also be used for more accurate resource estimates.







\begin{figure}[ht]
\centering
\includegraphics[width=1.0\textwidth]{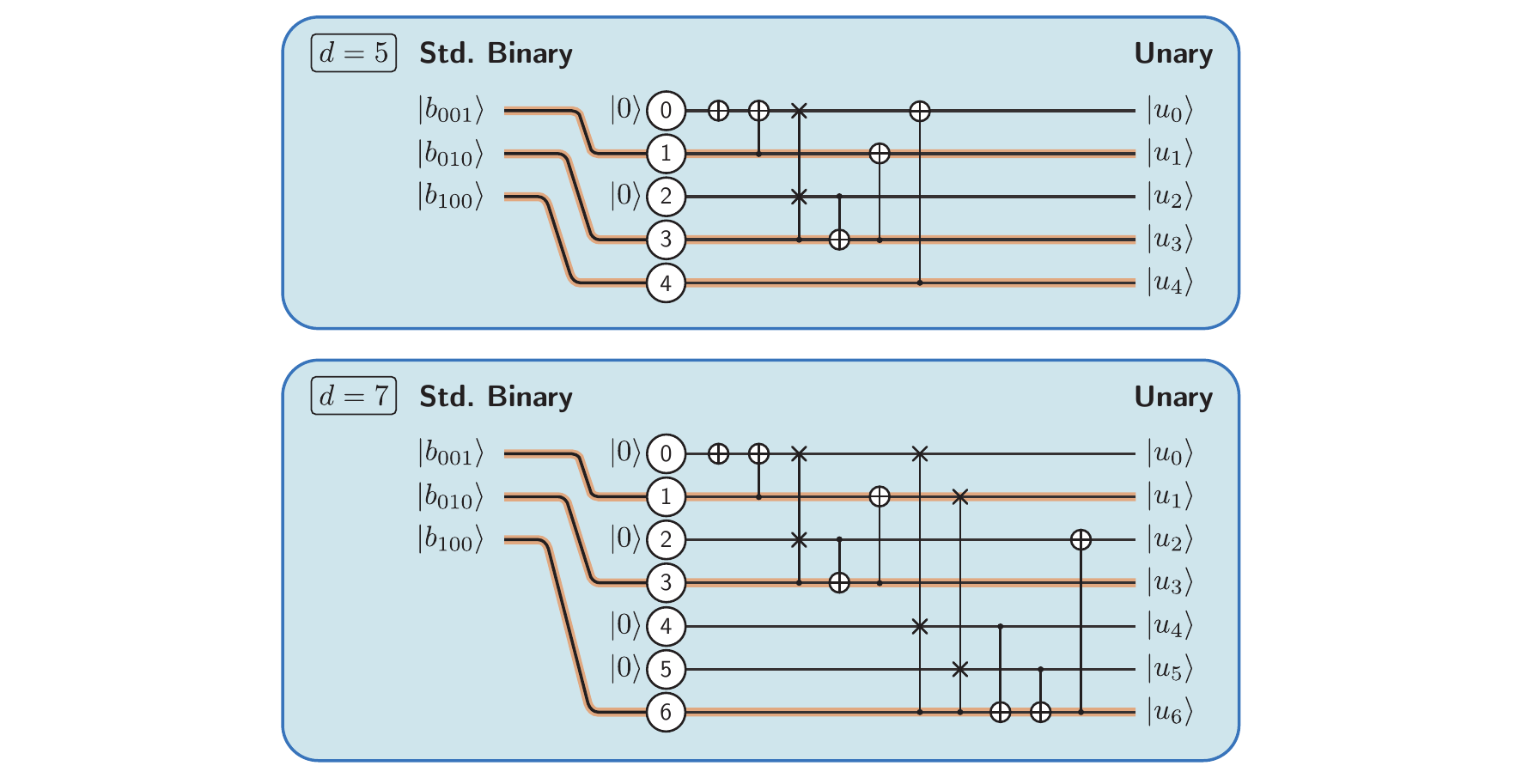}
\caption{Standard binary to unary conversion circuits for $d=5$ and $d=7$.}
\label{fig:conv_sb_unary_d5_d7}
\end{figure}

\section{Circuit optimization}



The number of required two-qubit entangling gates (CNOT gates) per Suzuki-Trotter step was determined as follows. Each $d$-level operator was converted to a Pauli operator using the procedure of the main text. Collecting coefficients and cancelling Pauli terms was done using the \texttt{QubitOperator} class of the python software packages FermiLib and OpenFermion \cite{openfermion}. Pauli terms were listed in a pseudo-alphabetical order, such that all terms containing $\sx^{(0)}$ are listed first, followed by terms beginning in $\sy^{(0)}$, $\sz^{(0)}$, $\sx^{(1)}$, $\sy^{(1)}$, etc. Quantum circuits were first synthesized by converting each Pauli term to a CNOT ladder, in the order described. One can see by inspection that circuit reductions will often occur. Pairs of CNOT, Hadamard, or $iR_{x+y}(\pi)$ gates, which appear after placing CNOT ladders next to each other, can be eliminated as they cancel to unity. 

The circuit optimization was performed as follows. In the optimizer, every gate is represented in a data structure that contains its attributes, \textit{e.g.} its inverse gate, its commutation properties with other gates, and whether it is a rotation gate. For each gate, the optimizer looks for an opportunity to cancel it with its inverse or merge it with another rotation gate by commuting it as far forwards and backwards as possible. The optimizer also looks for a limited set of commonly occurring patterns that allow for gate reductions. For the circuits used in this study, this pattern-searching allows us to reduce some sets of three CNOT gates to two: any gate sequence CNOT($i_0$,$i_1$) $\times$ CNOT($i_1$,$i_2$) $\times$ CNOT($i_0$,$i_1$) is changed to the equivalent CNOT($i_0$,$i_2$) $\times$ CNOT($i_1$,$i_2$). Merging and cancelling gates, as well as this pattern-searching, are performed for several passes until the circuit converges.

The choice of ordering for the Pauli terms affects the gate counts, as different orderings of CNOT ladders affect the presence of particular pairs of eliminable gates. Finding the absolute optimal ordering is a combinatorially hard problem and is not the focus of this work \cite{botea18}. However, to test the quality of the default ordering, we took the encoded bosonic $\hat q$ operator for the unary, gray, and standard binary codes, testing $>$900 random orderings for each. We found that the default ordering was superior to every random ordering. This result is intuitive, as orderings that match similar Pauli strings side-by-side will lead to more pairs of adjacent CNOT gates and adjacent single-qubit gates.

\section{SB to unary conversion: General case}

Here we discuss how to modify the SB-to-unary circuit shown in the main text when $d$ is not a power of two. First, we shift the most significant SB qubit up to qubit \textbf{$d-1$} rather than keeping it at qubit \textbf{$2^K -1$}, where $K$ is the number of qubits required by the standard binary encoding. Second, we change the control qubit of SWAP and CNOT gates from \textbf{$2^K-1$} to \textbf{$d-1$}, and remove all unnecessary qubits. Third, the target for the final CNOT is shifted up by the number of qubits that was removed. For instance, using $d=5$, the binary base line is shifted from \textbf{7} to \textbf{4}. We then remove qubits \textbf{4} through \textbf{6} and any associated gates. All SWAP and CNOT controlled by \textbf{7} become controlled by \textbf{4}. 
Finally, the last CNOT is placed with control \textbf{4} and target \textbf{0}.
The latter target has index $d-1-2^{K-1}$. Examples for $d=5$ and 7 are given in Supplementary Figure \ref{fig:conv_sb_unary_d5_d7}. Algorithm 1 in the main text correctly implements these cases; the examples discussed here are shown only as a pedagogical aid.

\section{SB to BU$^{SB}_{g=3}$ conversion}
\label{sec:firstquant}

\begin{figure}[ht]
\centering
\includegraphics[width=1.0\textwidth]{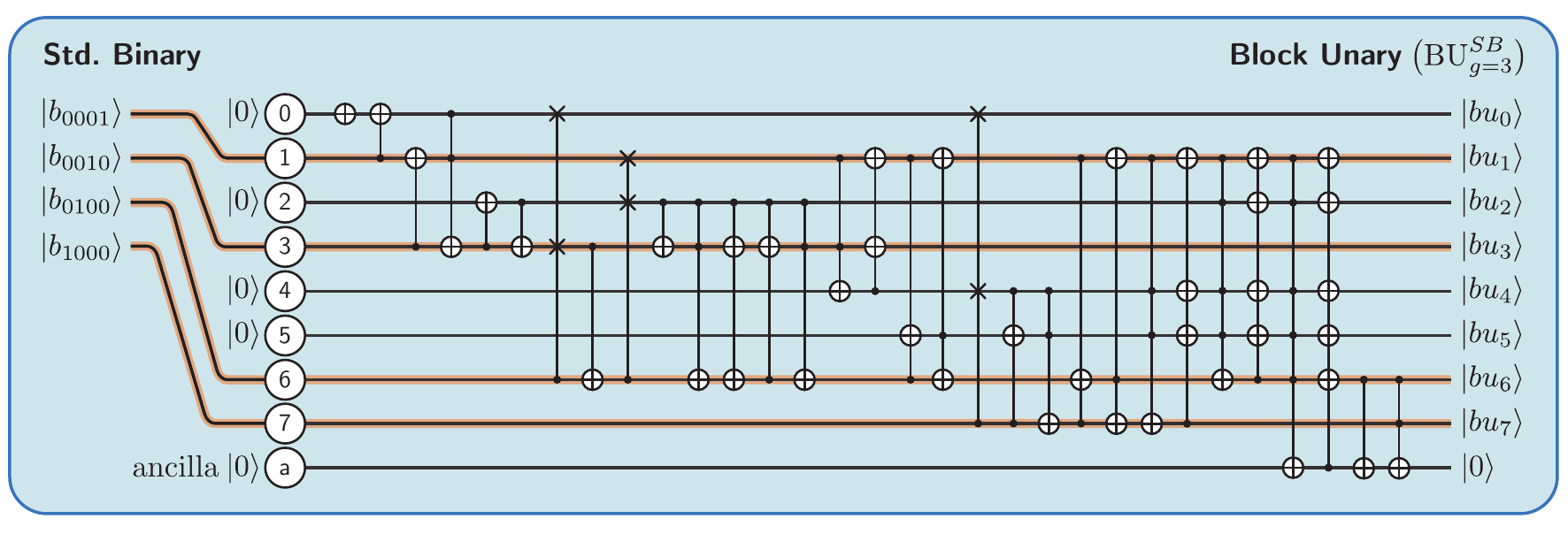}
\caption{Quantum circuit for standard binary to block unary (BU$^{SB}_{g=3}$) conversion.}
\label{fig:binary-to-block-unary-circuit}
\end{figure}

To show that one may in principle convert between all encodings, we provide a circuit for converting between SB and BU$^{SB}_{g=3}$ with $d=12$. We do not write down the general case for this circuit, because the rare hardware budgets for which BU might be useful would likely be cases for which the hardware's gate count constraints would not allow for the cost of conversion.

\section{First quantization}
\label{sec:firstquant}

First quantized operators can be used to represent any bosonic degree of freedom because first quantized QHO operators obey bosonic algebra \cite{macridin18a,macridin18b}. The utility of this approach comes partly from the fact that one can qssuickly move between the position and momentum bases by using the quantum Fourier transform, using a first quantized momentum operator $\tilde P_{\textrm{FQ}}$ that is the direct analog of $\tilde X_{\textrm{FQ}}$. For instance, consider implementing one Trotter step of an operator proportional to the approximate quantum harmonic oscillator Hamiltonian

\begin{equation}\label{eq:ham_qho_firstquant}
\tilde H_{\textrm{QHO}} \propto \tilde X_{\textrm{FQ}}^2 + \tilde P_{\textrm{FQ}}^2.
\end{equation}

One may first exponentiate $\tilde X^2$, then apply the quantum fourier transform (QFT) to move to the momentum basis, then apply $\tilde P^2$, and finally use QFT$^{-1}$ to move back to the position basis. This allows for all the operations to be done in the diagonal basis except for QFT, which takes just $O(N_q^2) \sim O(\log^2 d)$ operations. The Trotterization of $\tilde X_{\textrm{FQ}}$ and $\tilde P_{\textrm{FQ}}$ requires only $\log_2 d$ single-qubit rotations and zero two-qubit gates, as explained in the main text, although exponentiating their squares does require entangling gates.

However, a significant computational hurdle to the first quantization method is initial state preparation. Preparing the ground state of the QHO, for example, requires that one prepares the state 
\begin{equation} \label{eq:arbstate}
|\psi\ra = \sum_{j=0}^{K-1} f(x_j) | x_j \ra
\end{equation}

where $f(x)$ would be a normalized Gaussian, the first Hermite-Gauss function. Macridin et al. discuss existing options for preparing such an arbitrary state \cite{macridin18b}, showing that a variational approach is more efficient than one might expect for preparing ground states on near- or medium-term quantum computers. We expect that most early simulations of bosonic degrees of freedom on qubit-based quantum computers will use the Fock-type (\textit{i.e.} second quantization) mappings that are the focus of the current work, since preparing a Fock state in second quantization is trivially done by flipping individual bits. However, there may be instances in which Fock states are not used, such as wave packet dynamics or other approaches, and it may be that there are other operator-based advantages in using the first quantization mapping.
Note that the procedure for mapping $\tilde X$ and $\tilde P$ to qubits is no different than mapping any other matrix representation, with $d$ replaced by $N_x$.



\section{Hamiltonian definitions}
\label{sec:hamdefs}

In this section we provide the Hamiltonians studied in this work and discuss their relevance.

\textbf{Bose-Hubbard model.}
The Bose-Hubbard model is a ubiquitous model in condensed matter physics, consisting of a set of bosonic modes that occupy discrete lattice sites. The phase diagram of the basic model consists of the Mott insulator and superfluid phases \cite{fisher89}. Adding additional terms such as increasing the interaction radius or introducing an effective magnetic field can give rise to more complex phases like a density wave phase and a supersolid phase \cite{bh_phases_1, bh_phases_2, bh_phases_3}. The model has been experimentally realized in ultracold atomic lattices \cite{bloch_review}. Recent experiments \cite{supersolid_1, supersolid_2, supersolid_3} report evidence for the creation of one of the more exotic predicted phases, a supersolid, a state of matter where an ordered material can flow without friction.

Simulating Bose-Hubbard models on a quantum computer will help elucidate the properties of these phases and possibly lead to the discovery of previously unknown phases. Quantum simulation may be especially important for studying behavior near the boundaries of quantum phase transitions, a regime that is notoriously difficult to study numerically on classical computers \cite{huang17}.


Here, we consider the 1D Bose-Hubbard model with periodic boundary conditions, expressed as 
\begin{equation}\label{eq:hubb}
\hat H_{\textrm{BH}} = -t \sum_i^N (\hat a_i^\dag \hat a_{i+1} + h.c.) + \frac{U}{2}\sum_{i}^N \hat n_i(\hat n_i-1) - \mu \sum_i^N \hat n_i
\end{equation}

where $t$ is the hopping term, $U$ is the on-site site energy, and $\mu$ is the chemical potential. The parameters dictating the size of the problem are the number of levels $d$ per site as well as the number of particles $N$.




\textbf{Shifted one-dimensional QHO.} We consider this Hamiltonian because it is one of the simplest prototypes for encoding a bosonic system into qubits. The Hamiltonian is directly relevant to chemistry, where it is equivalent to calculating the Franck-Condon factors \cite{mcardle19,sawaya19} between two electronic transitions for a diatomic molecule. Franck-Condon factors determine the intensity of the transitions to vibrational eigenstates in the final electronic state.
The Hamiltonian is
\begin{equation}\label{eq:1dshift}
\hat H_{\textrm{1D Shift}} = \frac{\omega}{2}[(\hat q-\delta)^2+\hat p^2]
\end{equation}
where $\omega$ is the QHO frequency and $\delta$ is the displacement (related to the commonly used Huang-Rhys factor). The parameter $\delta$ indirectly determines how many levels $d$ will have appreciable intensity when considering overlaps with the unshifted QHO.





\textbf{Multidimensional Franck-Condon factors.} In the field of physical chemistry, in order to calculate an absorption spectrum, energy transfer characteristics, or other properties for molecules larger than two atoms, Franck-Condon profiles must include transitions to all possible vibrational Fock states of non-negligible intensity. Each vibrational normal mode in the final electronic state is a separate bosonic mode, and there are $M=3N-6(5)$ such modes in nonlinear (linear) molecules. One defines $\vec q_s$ and $\vec p_s$ to be vectors of position and momentum operators, respectively, where $s$ is one of two electronic potential energy surfaces (PESs) $A$ or $B$. See references \cite{huh11thesis,huh15,sawaya19} for more details. The unitary $M$-by-$M$ Duschinsky matrix $\mathbf S$ transforms between the bases of the two PESs such that 

\begin{equation}
\begin{split}
\vec q_B = \mathbf{\Omega_B S \Omega_A^{-1}} \vec q_A + \vec\delta  \\
\vec p_B = \mathbf{\Omega_B^{-1} S \Omega_A} \vec p_A
\end{split}
\end{equation}

where $\mathbf{\Omega}_s = diag( [\omega_{s1},...,\omega_{sM}] )^{\half}$ and $\{\omega_{si}\}$ are vibrational modes of PESs.
The Hamiltonian to simulate is
\begin{equation}\label{eq:fc}
\hat H_{\textrm{FC}} = \frac{1}{2} \sum_j^M \omega_{Bj} (q_{Bj}^2 + p_{Bj}^2)
\end{equation}

In the multidimensional case, all four parameters $\vec \delta$, $\mathbf S$, $\mathbf{\Omega_A}$, and $\mathbf{\Omega_B}$ indirectly determine the bosonic truncation in each mode, out of which the dimensionless displacement $\vec \delta$ usually has the greatest effect. Depending on these parameters, the required $d$ in a given mode can be anywhere from 1 in the case of negligible shifts, to as high as 70 in molecules such as nitrogen dioxide \cite{sawaya19}. Though we consider the harmonic case here, it is the anharmonic Hamiltonians, where higher-order Taylor series terms are added to \eqref{eq:fc}, where a quantum computer will likely first outperform a classical one for small molecules \cite{sawaya19,luis06,huh10,meier15,petrenko17}.

For real molecules, the Duschinsky matrix $\mathbf{S}$ tends to be sparse. In our resource count simulations, we assume that each row of $\mathbf{S}$ has a constant number of non-zero entries per row, $k$. We set $k=4$ for the simulations discussed here. We additionally assume that each vibrational mode is modelled with the same number of levels $d$, though for most real molecules one would use a different $d$ for each mode. We make the approximations in order to compare typical constant-factor savings between different mappings, independent of system size.








\textbf{
Spin-$s$ Heisenberg model.
}
Quantum Heisenberg models, commonly used to study theoretical questions in condensed matter physics and to model magnetism in real solids, consist of quantum spins placed on a lattice. Though most work has focused on spin-$\half$ particles, some problems require the use of higher-spin systems. Interpreting nuclear magnetic resonance experiments \cite{Levitt2008} can require modeling higher-spin systems. Notably, a Heisenberg model simulation using a high spin of spin-$\frac{7}{2}$ has been recently reported \cite{lora_serrano_16}, where the magnetic properties of the material GdRhIn$_5$ are studied. There are many stable isotopes up with nuclear spin $I=\frac{7}{2},4,\frac{9}{2}$ \cite{stone05}, and with unstable isotopes reaching still higher $I$.

For this work, we consider a one-dimensional spin-$s$ Heisenberg model with ferromagnetic $\Sz \Sz$ interactions and a magnetic field in the transverse direction, described by Hamiltonian

\begin{equation}\label{eq:heisenberg}
\hat H_{\textrm{Heis}} = -J \sum_{\langle i,j \rangle} \Sz^{(i)} \Sz^{(j)} - gJ \sum_i  \Sx^{(i)} \; ,
\end{equation}
\textbf{Boson sampling.} There is strong evidence that simulating multi-mode linear optical devices is hard. The problem has been formalized and can be stated as follows: consider a linear optical device with $M$ modes and in which $N$ single photons are injected in the first $N\leq M$ modes. The photons are detected in the output modes with occupation numbers that vary from repetition to repetition of the experiment. The task is to sample from this output distribution and, under the assumption that $N \ll M$ and that the linear optical device has been chosen at random (see \cite{aaronson14} for details), together with some common theoretical conjectures, it has been shown that such a task is classically hard to perform. 
Such devices are composed of two kinds of optical elements: single-mode phase shifters and two-mode beam splitters. Here we consider the possibility of implementing the same dynamics not with optical systems, but using qubits. This approach may find utility for example in testing computational complexity conjectures, if advanced qubit-based quantum computers come to fruition before advanced dedicated bosonic devices. A recent work discussed mappings to qubits for simulating quantum optics \cite{sabin19}.

In contrast to the other examples in this section, the dynamics of boson sampling is described in a stroboscopic way with layer-by-layer unitary operations, instead of being described by the time evolution of a large Hamiltonian. Nevertheless, the results of the previous sections can be readily applied when we describe the unitary operations in terms of their generators, in practice providing the effective Hamiltonians governing the evolution in each optical element. For single-mode phase shifter one has:
\begin{equation}
    \hat h_{\textrm{p.s.}} = \hat a_i^\dagger \hat a_i
\end{equation}
while for two-mode beam splitter one has:
\begin{equation}
    \hat h_{\textrm{b.s.}} = \hat a_i^\dagger \hat a_j + \hat a_j^\dagger \hat a_i
\end{equation}
with indices $i,j$ labelling the different optical modes. 

Note that this implementation is different from the previous Hamiltonian simulation examples, because the goal is to implement
\begin{equation}
\hat U_{\textrm{Boson Sampling}}
    = \exp(-i \theta_R \hat h_R) \exp(- i \theta_{R-1} \hat h_{R-1}) \cdots \exp(-i \theta_1 \hat h_1)
\end{equation}
where $\hat h_1,\cdots,\hat h_R$ are the ordered list of effective Hamiltonians for each gate considered. Effectively, we produce quantum circuits that, after encoding the problem to qubits, performs a first-order Trotterization on each separate unitary. 




\section{Full Composite Hamiltonian Results}
\label{sec:composite_full}

\begin{figure}[]
\centering
\includegraphics[width=0.85\textwidth]{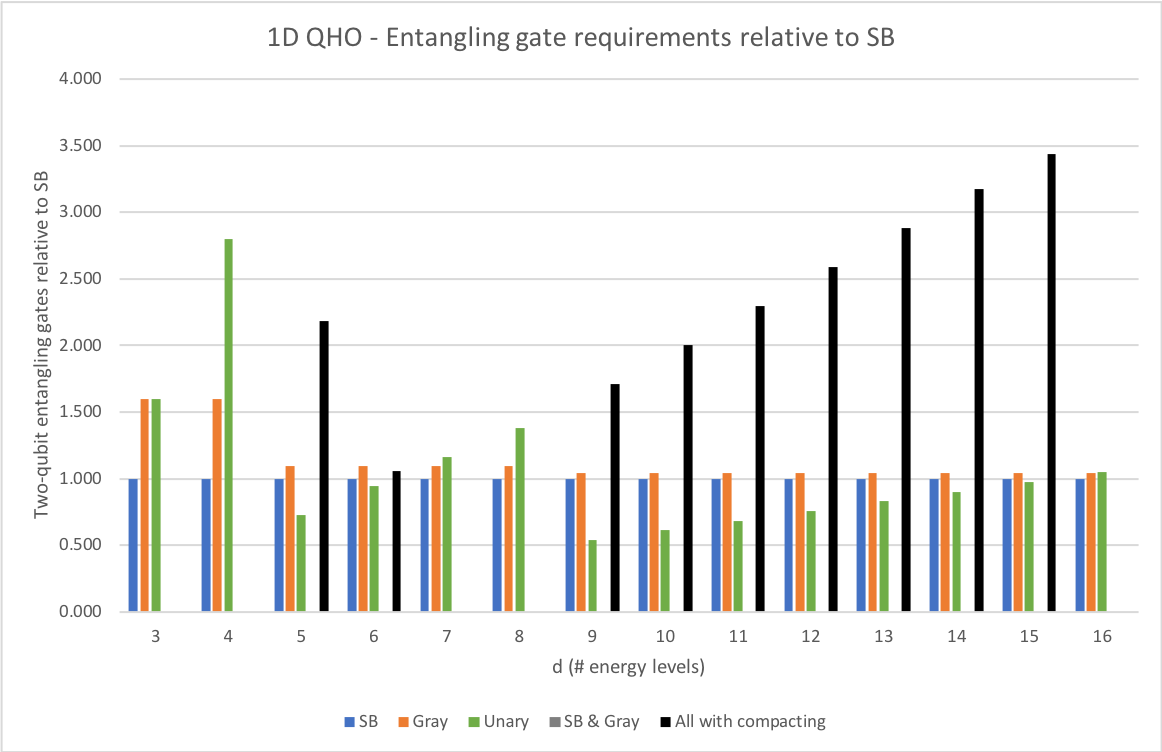}
\caption{Two-qubit entangling gate resource requirements, relative to SB, for the one-dimensional QHO. In order from $d$=3 through 16, the encoding \textit{scenarios} are respectively \textit{A, A, D, D, A, A, D, D, D, D, D, D, D, A}.}
\label{fig:fullhist_qho1d}
\end{figure}

\begin{figure}[]
\centering
\includegraphics[width=0.85\textwidth]{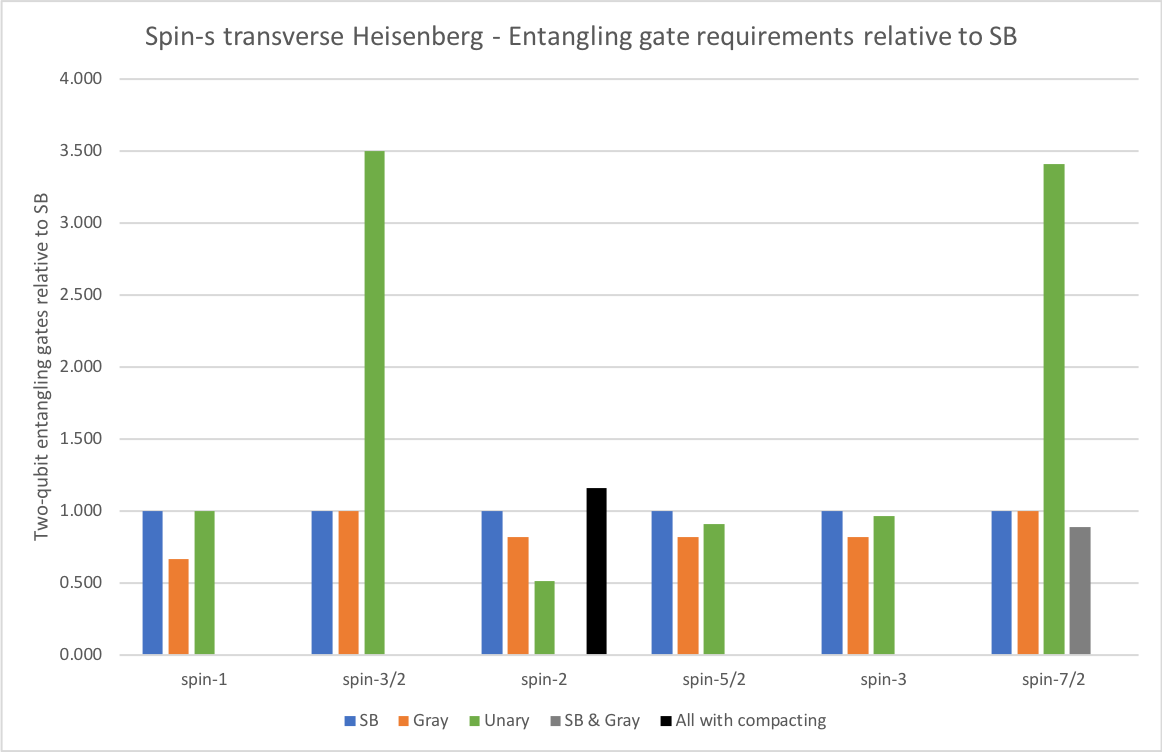}
\caption{Two-qubit entangling gate resource requirements, relative to SB, for a spin-$s$ transverse Heisenberg model with an arbitrary number of particles $N$. In order from $s$=1 through $\frac{7}{2}$, the encoding \textit{scenarios} are respectively \textit{A, A, D, A, A, B}.}
\label{fig:fullhist_heis}
\end{figure}

\begin{figure}[]
\centering
\includegraphics[width=0.85\textwidth]{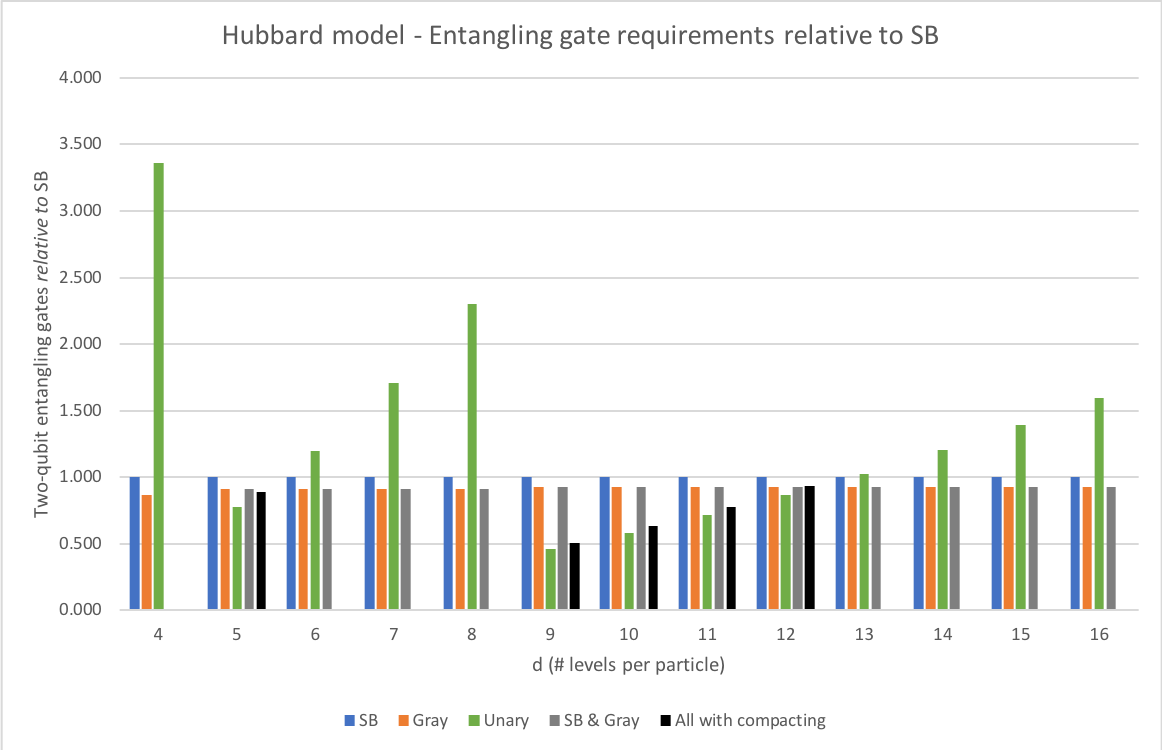}
\caption{Two-qubit entangling gate resource requirements, relative to SB, for a Bose-Hubbard model with an arbitrary number of particles $N$. In order from $d$=4 through 16, the encoding \textit{scenarios} are respectively \textit{A, C, B, B, B, C, C, C, D, B, B, B, B}.}
\label{fig:fullhist_hubb}
\end{figure}

\begin{figure}[]
\centering
\includegraphics[width=0.85\textwidth]{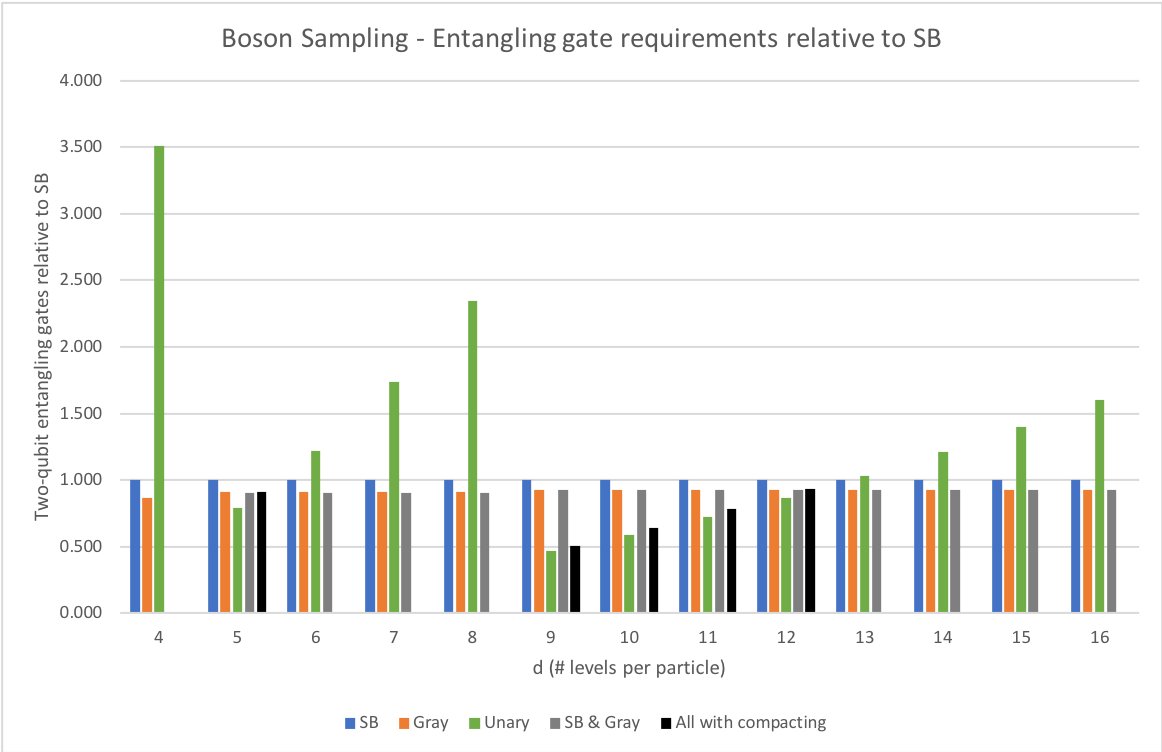}
\caption{Two-qubit entangling gate resource requirements, relative to SB, for a digital quantum simulation of boson sampling, with an arbitrary number of photonic modes. In order from $d$=4 through 16, where $d-1$ is the maximum number of photons per optical mode, the encoding \textit{scenarios} are respectively \textit{A, D, B, B, B, C, C, C, D, B, B, B, B}.}
\label{fig:fullhist_bossamp}
\end{figure}

\begin{figure}[]
\centering
\includegraphics[width=0.85\textwidth]{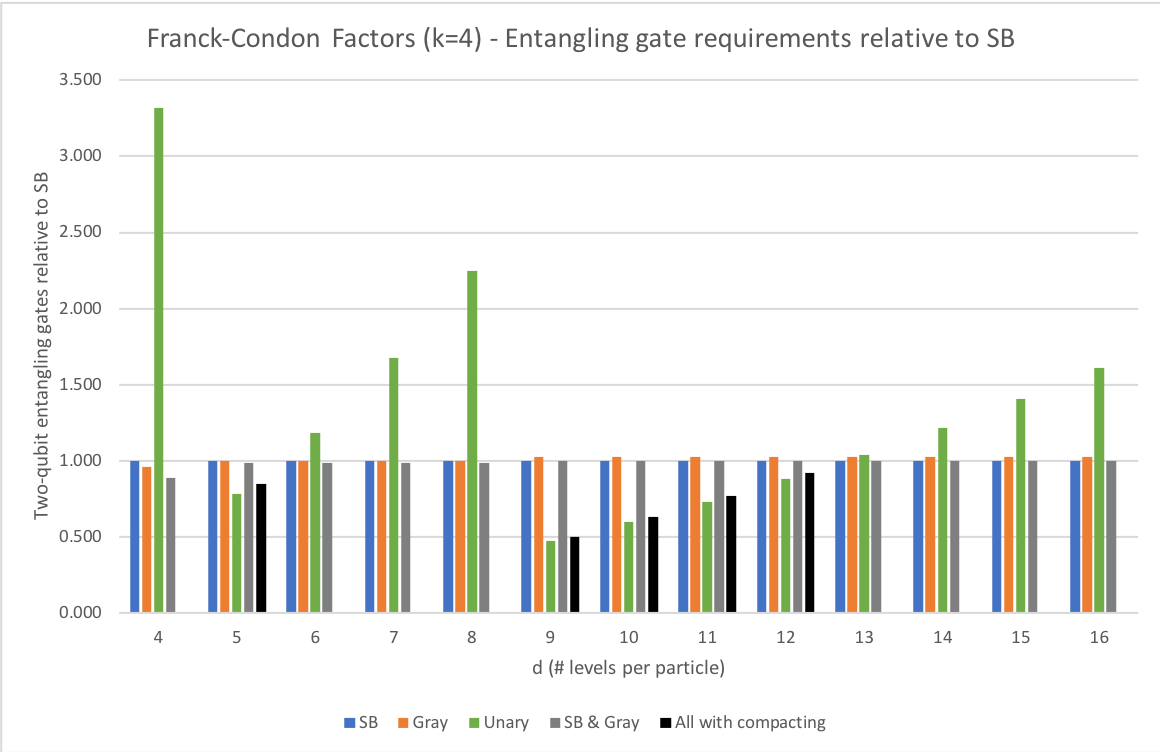}
\caption{Two-qubit entangling gate resource requirements, relative to SB, for implementing a Franck-Condon Hamiltonian of a molecule of arbitrary size, assuming k=4. In order from $d$=4 through 16, the encoding \textit{scenarios} are respectively \textit{B, C, B, B, B, C, C, C, C, B, B, B, B}.}
\label{fig:fullhist_fc}
\end{figure}

Supplementary Figures \ref{fig:fullhist_qho1d} -- \ref{fig:fullhist_fc} show two-qubit entangling gate resource requirements relative to SB, respectively for the one-dimensional QHO, spin-$s$ transverse Heisenberg model, Bose-Hubbard model, boson sampling problem, and Franck-Condon problem. The latter four cases correspond to an arbitrary number of particles or modes. The bosonic Hamiltonians are plotted up to $d=16$ and the Heisenberg model up to $s=\frac{7}{2}$. Scenarios \textit{A} through \textit{D} (defined in the main text), to which each combination $d$ and problem class belong, are given in the captions. As stated in the main text, results for converting between Gray and SB are omitted when they are no better than Gray-only and SB-only. Additionally, results for expanding and re-compacting (``All with compacting'') are omitted if unary-only is not better than all the compact schemes. Note that many of the ``SB \& Gray'' results included in the plots are only marginally ($<$ 1\%) superior to the SB-only and Gray-only results.

\clearpage

\clearpage

\bibliographystyle{alpha}
\bibliography{refs}